\UseRawInputEncoding
\documentclass{article}

\usepackage{amsmath, amsthm, amssymb, geometry, mathtools, epsfig}
\usepackage[colorlinks,citecolor=blue,linkcolor=red]{hyperref}

\def\G{\Gamma}
\def\no{\nonumber}

\def\dis{\displaystyle}
\def\le{\left(}
\def\ri{\right)}

\def\res{\mathop{{\rm Res}}\limits}

\begin{document}

\begin{titlepage}
\vskip 1cm
\begin{center}
{\Large \bf Analytical solution to DGLAP integro-differential equation \\ 
\vskip 3mm
in a simple toy-model with a fixed gauge coupling }   \\ 
\vskip 5mm  
Gustavo \'Alvarez $^{(a,b)},$
Gorazd Cveti\v{c}  $^{(c)},$
Bernd A. Kniehl $^{(a)},$ \\
\vskip 2mm
Igor Kondrashuk $^{(d)},$ 
Ivan Parra-Ferrada $^{(e)}$ 
\vskip 5mm  
{\it  (a) II. Institut f\"ur Theoretische Physik, Universit\"at Hamburg,  \\ Luruper Chaussee 149,  22761 Hamburg, Germany} \\ 
{\it  (b) Departamento de F\'isica, Universidad de Concepci\'on, Casilla 160-C, Concepci\'on, Chile} \\
{\it  (c) Departamento de F\'isica,   Universidad T\'ecnica Federico  Santa Mar\'ia, \\  Casilla 110-V, Valpara\'iso, Chile} \\
{\it  (d) Grupo de Matem\'atica Aplicada {\rm \&} Grupo de F\'isica de Altas Energ\'ias  {\rm \&}  Centro de Ciencias \\ 
         Exactas   {\rm \&}  Departamento de Ciencias B\'asicas, Universidad del B\'io-B\'io, \\ 
         Campus Fernando May, Av. Andres Bello 720, Casilla 447, \\ 
         Chill\'an, Chile} \\
{\it  (e) Instituto de Matem\'atica y F\'isica, Universidad de Talca, \\ 2 Norte 685, Casilla 721, Talca, Chile } 
\end{center}

\begin{abstract}
We consider a simple model for QCD dynamics in which DGLAP integro-differential equation may be solved analytically. 
This is a gauge model which possesses dominant evolution of gauge boson (gluon) distribution and in which 
the gauge coupling does not run. This may be ${\cal N} =4$ supersymmetric gauge theory with softly 
broken supersymmetry, other finite supersymmetric gauge theory with lower level of supersymmetry, 
or topological Chern-Simons field theories. We maintain only one term  in the splitting function 
of unintegrated gluon distribution and solve DGLAP analytically for this simplified splitting function. 
The solution is found by use of the Cauchy integral formula. The solution restricts form of the unintegrated 
gluon distribution as function of momentum transfer and of Bjorken $x$. Then we consider an almost realistic 
splitting function of unintegrated gluon distribution as an input to DGLAP equation and solve it by the same method 
which we have developed to solve DGLAP equation for the toy-model. We study a result obtained for the realistic gluon distribution and 
find a singular Bessel-like behaviour in the vicinity of the point $x=0$ and a smooth behaviour in the vicinity of the point
$x=1.$
\end{abstract}
\end{titlepage}

\section{Introduction}

DGLAP equation is a renormalization group equation (RGE) for the integrated parton distributions.  It has been written initially for QED 
\cite{Gribov:1972ri,Gribov:1972rt,Lipatov:1974qm} 
in an integro-differential form. BFKL equation appears as a result of generalization of the Regge theory of scattering from quantum mechanics to 
QCD \cite{Lipatov:1976zz,Fadin:1975cb,Kuraev:1976ge,Kuraev:1977fs,Balitsky:1978ic}. 
In Refs. \cite{Dokshitzer:1977sg,Altarelli:1977zs} the DGLAP equation was written as a RGE for the integrated parton distributions in QCD.  
Dokshitzer \cite{Dokshitzer:1977sg} wrote this equation in an integro-differential form based on Gribov and Lipatov results in QED 
\cite{Gribov:1972ri,Gribov:1972rt,Lipatov:1974qm} and also 
Bethe-Salpeter technique used earlier in  the BFKL equation was applied.

The BFKL equation is an optic theorem written down for the amplitude of scattering of two particles $A(s,t)$ in the Regge limit. The amplitude 
$A(s,t)$ may be obtained from the four-point Green function of the reggeized gluons after integrating the part of the external momenta with the 
impact factors. The optic theorem may be mapped to an integro-differential equation (IDE) for this four-point Green function (and for the amplitude 
$A(s,t)$ in the Regge limit), in which the derivative is taken  with respect to $\ln s.$  The four-point Green function depends on 
the variable $t$ too. The BFKL IDE is written for unintegrated gluon distributions.

The DGLAP equation is another IDE  in which the derivative is taken with respect to variable $u = Q^2/\mu^2,$ where $Q^2$ is the momentum transfer in the $t$-channel of 
the two  particles in two-particles scattering process and the kernel of this IDE depends  on the variables $t = -Q^2$  and  $x.$  The DGLAP IDE may be considered as the  RGE for integrated parton distributions 
and is valid for large $t = -Q^2$ and large $x$ in order to be in the framework of the perturbation theory.  The DGLAP IDE
may be written as a matrix differential equation  in which the derivative is taken with respect to variable $u$ too. 
This matrix differential equation is written for the Mellin moment $G(N,u)$  of integrated gluon distribution  $G(x,u)$ and $\Sigma(N,u)$ of integrated singlet distribution  $\Sigma(x,u).$
The procedure of the integral transformation to the Mellin moments suggests that $N$ is a complex variable.  In Refs. 
\cite{Ball:1999sh,Altarelli:1999vw,Altarelli:2000dw,Altarelli:2000mh,Altarelli:2001ji,Altarelli:2003hk,Altarelli:2005ni,Ball:2005mj} an approximation 
of DGLAP matrix differential equation has been considered in which the Mellin moment $\Sigma(N,u)$ of the integrated singlet distribution was discarded and  
the Mellin moment of integrated gluon distribution $G(x,u)$ was considered only. In these articles  the saddle 
point method was used  to find an approximate  solution of DGLAP equation for the Mellin moments.

The BFKL and DGLAP equations are unstable under radiative corrections in the different regimes. For example, DGLAP splitting functions 
are unstable at small $x$ and BFKL kernel is unstable at large momentum transfer $Q^2$. 
In Refs. \cite{Ball:1999sh,Altarelli:1999vw,Altarelli:2000dw,Altarelli:2000mh,Altarelli:2001ji,Altarelli:2003hk,Altarelli:2005ni,Ball:2005mj}
it has been shown that the both IDEs may be considered together on the same footing and the problem of their stability has been treated.

In Refs. \cite{Kotikov:2000pm,Kotikov:2002ab} the relation between DGLAP and BFKL equations were studied from a different point of view. 
In ${\cal N} = 4$ supersymmetric Yang-Mills theory due to the vanishing of $\beta$ function the DGLAP splitting functions are stable even 
for small $x$  and the matrix of anomalous dimensions may be obtained from the BFKL equation  \cite{Kotikov:2002ab}.
In ${\cal N} =4$ supersymmetric Yang-Mills theory the matrix of anomalous dimensions was obtained explicitly from the BFKL eigenvalues 
\cite{Kotikov:2002ab} without making any conjecture about the form of the splitting functions $P$ which stand in the integral kernels in DGLAP. 
A possibility to obtain a matrix of anomalous dimensions from the BFKL eigenvalues in nonsupersymmetric QCD was considered   in
\cite{Ball:1999sh,Altarelli:1999vw,Altarelli:2000dw,Altarelli:2000mh,Altarelli:2001ji,Altarelli:2003hk,Altarelli:2005ni,Ball:2005mj}.

\section{Integral Transforms}

In this section we collect necessary formulas of various integral transforms which we will use in all the paper.   

\subsection{Mellin transform}

We define Mellin transform as 
\begin{eqnarray} \label{MT}
MT[f(x),x](z) = \int_0^{\infty} x^{z-1}f(x)~dx, 
\end{eqnarray}
in which the arguments in the brackets on the l.h.s. stand for the transforming function $f(x)$  and the integration variable $x$ of 
this integral transformation. The inverse Mellin transformation is 
\begin{eqnarray} \label{MT Cauchy} 
f(x) = \frac{1}{2\pi i}\int_{c-i\infty}^{c + i\infty}x^{-z} MT[f(x),x](z) ~dz, ~~~ x \in [0,\infty[.
\end{eqnarray}
The position point $c$ of the  vertical line  of the integration contour 
in the complex plane must be in the vertical strip $c_1 < c < c_2,$ the borders of the strip are defined by the condition that two integrals 
\begin{eqnarray} 
\int_0^{1} x^{c_1-1}f(x)~dx  &  {\rm and } &  \int_1^{\infty} x^{c_2-1}f(x)~dx
\end{eqnarray}
must be finite. 
This means that
\begin{eqnarray}
|f(x)| < 1/x^{c_1} \quad {\rm when} \; x \to +0, 
\qquad
|f(x)|< 1/x^{c_2} \quad {\rm when} \; x \to + \infty .
\end{eqnarray}
Should the contour in  Eq.(\ref{MT Cauchy}) be closed to the left complex infinity or to the right complex infinity  
depends on the explicit asymptotic behaviour of the Mellin transform  $MT[f(x),x](z)$ at the complex infinity. 
We close to the left if the left complex infinity does not contribute 
and we close to the right if the right complex infinity does not contribute \footnote{In comparison, in the Mellin-Barnes transformation 
we choose to which infinity the contour should be closed by taking into account 
the absolute value of $x$ in  (\ref{MT Cauchy}) because the MB transform has already an established structure 
in a form of fractions of the Euler $\G$ functions. However, MB transformation is only a particular case of Mellin transformation.}.
Under this condition the original function $f(x)$ may be reproduced  via calculation of the residues by Cauchy formula. 

One of the simplest examples of the Mellin transformation is
\begin{eqnarray} 
\G(z) = \int_0^{\infty} e^{-x}x^{z-1}~dx & {\rm and }  &   e^{-x} = \frac{1}{2\pi i}\int_{c-i\infty}^{c + i\infty}x^{-z} \G(z) ~dz.
\end{eqnarray}
The contour in the complex plane is the vertical line with ${\rm Re}~z = c$ is in the strip $0 < c < A,$ where $A$ is a real 
and positive number, the contour must be closed to the left infinity.
 
We may write many parameters (for example,  other complex variables), $\overrightarrow{\alpha} = \le \alpha_1,\dots,\alpha_n\ri$  on which 
the function $f$ may depend,
\begin{eqnarray}  
MT[f(x,\overrightarrow{\alpha}),x](z) \equiv M[f(x,\alpha_1,\dots,\alpha_n),x](z) 
= \int_0^{\infty} x^{z-1}f(x,\alpha_1,\dots,\alpha_n)~dx  \no \\ 
\equiv   \int_0^{\infty} x^{z-1}f(x,\overrightarrow{\alpha})~dx. 
\end{eqnarray}

The integral on the r.h.s. of Eq.(\ref{MT}) may be seen as a sum of two integrals 
\begin{eqnarray} \label{MT-LT} 
\int_0^{\infty} x^{z-1}f(x)~dx = \int_0^{1} x^{z-1}f(x)~dx + \int_1^{\infty} x^{z-1}f(x)~dx \\
= \int_{-\infty}^{0} e^{tz}f(e^t)~dt + \int_{0}^{\infty} e^{tz}f(e^t)~dt =  \int_{-\infty}^{\infty} e^{tz}f(e^t)~dt. \no
\end{eqnarray}

\subsection{Laplace transform}

Representation (\ref{MT-LT}) of the Mellin transformation is closely related to the Laplace transformation. 
We define Laplace transform of function $f(x)$ as 
\footnote{We note that in most of bibliographical references the Laplace transformation is defined differently, as  
$L[f(x),x](z) = \int_{-\infty}^{\infty} e^{-xz}f(x)~dx$ }
\begin{eqnarray}  \label{LT}
L[f(x),x](z) = \int_{0}^{\infty} e^{-xz}f(x)~dx. 
\end{eqnarray}
This transformation is defined only for the functions that have restricted exponential growth $a,$ that is $f(x) < Ae^{a x},$ $A$ 
is a real positive, in the right complex half-plane ${\rm Re}~z > a.$ In this case the inverse transformation is  
\begin{eqnarray}  \label{LT Cauchy} 
f(x) = \frac{1}{2\pi i}\int_{a+\delta-i\infty}^{a+\delta+i\infty}e^{xz} L[f(x),x](z) ~dz, 
\end{eqnarray}  
where ${\rm Re} (z) = a + \delta$ and $\delta \to +0$. This means that the vertical line of the integration in the complex plane passes slightly to the right of the point $a$.
To show the compatibility explicitly,  we perform subsequent transformations and obtain identity 
\begin{eqnarray} 
L[f(x),x](z) = \frac{1}{2\pi i}\int_{0}^{\infty} e^{-xz} ~dx \int_{a+\delta-i\infty}^{a+\delta+i\infty}e^{xu} L[f(x),x](u) ~du =  \no\\
\frac{1}{2\pi i}\int_{a+\delta-i\infty}^{a+\delta+i\infty}\frac{L[f(x),x](u)}{z-u}du =  L[f(x),x](z),
\end{eqnarray}
where ${\rm Re}~z > a +\delta > a.$  The contour is closed to the right complex infinity. We cannot close the contour to the left infinity 
since $L[f(x),x](z)$ has poles in the half-plane to the left from the vertical line which crosses the real axis at the point $a+\delta.$ The inverse Laplace transformation 
can be checked as 
\begin{eqnarray}
f(x) = \frac{1}{2\pi i}  \int_{a+\delta-i\infty}^{a+\delta+i\infty}e^{xz}  L[f(x),x](z) dz = \frac{1}{2\pi i}  
\int_{a+\delta-i\infty}^{a+\delta+i\infty}e^{xz} \int_{0}^{\infty} e^{-uz} f(u) ~du dz = \no\\
\int_{0}^{\infty} \delta(x-u) f(u) ~du  = f(x),
\end{eqnarray}
this is valid due to the following integral relation
\begin{eqnarray}
\frac{1}{2\pi i}  \int_{a+\delta-i\infty}^{a+\delta+i\infty}e^{(x-u)z}  dz = \frac{1}{2\pi }  \int_{-\infty}^{\infty}e^{(x-u)(a+\delta +i\tau)}  d\tau = 
\frac{e^{(x-u)(a + \delta)}}{2\pi }\int_{-\infty}^{\infty}e^{i(x-u)\tau}  d\tau = \no\\
e^{(x-u)(a + \delta)}\delta(x-u) = \delta(x-u).
\end{eqnarray}

\subsection{Mellin moments}

We define Mellin $z$-moment of function $f(x)$ as  
\begin{eqnarray} \label{MMT} 
M[f(x),x](z) =  \int_0^{1} x^{z-1}f(x)~dx,
\end{eqnarray}
$z$ is a complex variable.    To construct the inverse transformation, 
we need to rewrite (\ref{MMT}) in the form of the Laplace transformation (\ref{LT}) and then to use  (\ref{LT Cauchy}), 
\begin{eqnarray}  
L[f(x),x](z) = \int_{0}^{\infty} e^{-xz}f(x)~dx =  \int_{-\infty}^{0} e^{xz}f(-x)~dx = \int_{0}^{1} y^{z-1}f(-\ln y)~dy \no\\
\equiv \int_{0}^{1} y^{z-1}F(y)~dy \equiv M[F(y),y](z),
\end{eqnarray}
where we have introduced a new function $F(y) \equiv f(-\ln y).$  The Laplace transform for the function $f(x)$ appears to be
a Mellin moment for the function $F(y),$  
\begin{eqnarray} \label{MMT Cauchy}
 \dis{f(x) = \frac{1}{2\pi i}\int_{a+\delta-i\infty}^{a+\delta+i\infty}e^{xz} L[f(x),x](z) ~dz} \Rightarrow \no\\
 \dis{F(y) = f(-\ln y) = \frac{1}{2\pi i}\int_{a+\delta-i\infty}^{a+\delta+i\infty}y^{-z} L[f(x),x](z) ~dz} \no\\
 = \dis{\frac{1}{2\pi i}\int_{a+\delta-i\infty}^{a+\delta+i\infty}y^{-z} M[F(y),y](z) ~dz } 
\end{eqnarray}

Since the Laplace transform  $L[f(x),x](z)$  is defined in the domain  ${\rm Re}~z > a,$ where $a$ is an index of the exponential growth of 
the function $f(x),$ the Mellin moment   $M[F(y),y](z)$ is defined in the same domain because the power-like restriction on its growth
\begin{eqnarray}\label{Power-like growth-MMT}
F(y) < A/y^{a} 
\end{eqnarray}
comes from the restrictions on $f(x).$ In the inverse transformation   the contour passes vertically in the complex plane $z$
in the same position at ${\rm Re}~z = a + \delta$ as it does for the Laplace transformation (\ref{LT Cauchy}). Under this condition the  
$M[F(y),y](z)$ does not have poles in the complex half-plane to the right from this vertical line.  

The direct proof of the inverse transformation may be done as 
\begin{eqnarray} 
M[F(y),y](z) = \frac{1}{2\pi i}\int_{0}^{1} y^{z-1} ~dy \int_{a+\delta-i\infty}^{a+\delta+i\infty}y^{-u} M[F(y),y](u) ~du = \no \\
\frac{1}{2\pi i}\int_{a+\delta-i\infty}^{a+\delta+i\infty}\frac{M[F(y),y](u)}{z-u}du =  M[F(y),y](z),
\end{eqnarray}
and the inverse transformation may be proved as 
\begin{eqnarray}
F(y) = \frac{1}{2\pi i}  \int_{a+\delta-i\infty}^{a+\delta+i\infty}y^{-z}  M[F(y),y](z) dz 
= \frac{1}{2\pi i}  \int_{a+\delta-i\infty}^{a+\delta+i\infty}y^{-z} \int_{0}^{1} u^{z-1} F(u) ~du dz = \no \\
\int_{0}^{1} \delta\le\ln{\frac{u}{y}}\ri F(u) u^{-1}~du  = F(y),
\end{eqnarray}
where $0 < y < 1.$  Thus, the transformation (\ref{MMT Cauchy}) is inverse to transformation (\ref{MMT}) 
under the restriction for the power-like growth  (\ref{Power-like growth-MMT}).

The Mellin moments, Laplace transform and Mellin transform posses the same equation for the inverse transformation. However, they are related by 
complex diffeomorphisms.

\section{Description of Theoretical Setup}

Structure functions of nucleons may be measured in deep inelastic scattering processes. They are related to integrated parton distributions which have probabilistic interpretation.
There are two integro-differential equations (IDEs) for parton distributions, DGLAP equation \cite{Dokshitzer:1977sg,Altarelli:1977zs}  and 
BFKL equation \cite{Lipatov:1976zz,Fadin:1975cb,Kuraev:1976ge,Kuraev:1977fs,Balitsky:1978ic}, studied widely in many papers. 
Our paper is dedicated to a toy-model for evolution of integrated gluon distribution.  We present an analytical solution to DGLAP equation in this model.

\subsection{Evolution Equations}

Because we need to use the evolution equations along all the paper, we briefly review the main idea of the probabilistic interpretation of them 
along the line of Ref. \cite{Altarelli:1977zs}. The IDE in which participates the splitting function $P(x)$ is 
\begin{eqnarray} \label{DGLAP-1} 
u\frac{d}{du} f(x,u) =  \frac{\alpha(u)}{2\pi}\int_x^1\frac{dy}{y}~f(y,u)P\left(\frac{x}{y}\right).
\end{eqnarray}
We calculate Mellin $N$-moment of both the parts of this equation and obtain the relation 
\begin{eqnarray}
u\frac{d}{du} \int_0^{1}dx~x^{N-1}f(x,u)  =  \frac{\alpha(u)}{2\pi}\int_0^{1}dx~ x^{N-1}\int_x^1\frac{dy}{y}~f(y,u)P\left(\frac{x}{y}\right)  \no \\
=  \frac{\alpha(u)}{2\pi} \int_0^{1}dy~f(y,u)\frac{1}{y} \int_0^y dx ~x^{N-1} P\left(\frac{x}{y}\right) = \frac{\alpha(u)}{2\pi}\int_0^{1}dy~y^{N-1}f(y,u) \int_0^1 dx ~x^{N-1} P\left(x\right)  \no \\
= \frac{\alpha(u)}{2\pi} ~\gamma(N,\alpha(u)) ~M[f(y,u),y](N),
\end{eqnarray}
where we define $\gamma(N,\alpha(u))$ as
\begin{eqnarray}
\int_0^1 dx ~x^{N-1} P\left(x\right) = \gamma(N,\alpha(u)), ~~~ \gamma(1,\alpha(u))  = 1. \no \\
\end{eqnarray}
Thus, the RGE 
\begin{eqnarray} \label{RGEq} 
u \frac{d}{d u} M[f(x,u),x](N)  = \frac{\alpha(u)}{2\pi}~\gamma(N,\alpha(u))~M[f(x,u),x](N).
\end{eqnarray}
may be re-written in the form of IDE (\ref{DGLAP-1}). A complex variable\footnote{Letter $N$ is used in order to agree with the notation of Refs. 
\cite{Ball:1999sh,Altarelli:1999vw,Altarelli:2000dw,Altarelli:2000mh,Altarelli:2001ji,Altarelli:2003hk,Altarelli:2005ni,Ball:2005mj}.}
$N$ appears in the Mellin moment $M[f(x,u),x](N)$ of function $f(x,u).$

\subsection{Parton distributions}

In the realistic QCD dynamics when there are integrated quark distributions  $q_i(x,u)$ of different flavors $i$ and there is integrated gluon distribution $G(x,u),$ the system evolves according to IDEs  
given in \cite{Altarelli:1977zs}, 
\begin{eqnarray}
\label{differences} u\frac{d}{du}\Delta_{ij}(x,u) &=& \frac{\alpha(u)}{2\pi}\int_x^1\frac{d y}{y}\Delta_{ij}(y,u)P_{qq}\left(\frac{x}{y}\right)\\
\label{SD} u\frac{d}{du} \Sigma (x,u) &=& \frac{\alpha(u)}{2\pi}\int_x^1\frac{dy}{y}\left[\Sigma(y,u)P_{qq}\left(\frac{x}{y}\right) + (2f)G(y,u)P_{qG}\left(\frac{x}{y}\right)\right]\\
\label{GD} u\frac{d}{du} G(x,u)  &=& \frac{\alpha(u)}{2\pi}\int_x^1\frac{d y}{y}\left[\Sigma(y,u)P_{Gq}\left(\frac{x}{y}\right)+ G(y,u)P_{GG}\left(\frac{x}{y}\right)\right] 
\end{eqnarray}
with $\Delta_{ij}(x,u) = q_i(x,u) - q_j(x,u)$ and $\Sigma(x,u)= \sum_i \left[q_i(x,u) + \overline{q}_i(x,u)\right],$  
where $\Sigma(x,u)$  is called integrated singlet distribution  and $\Delta_{ij}(x,u)$ are called non-singlet integrated quark distributions.  
Splitting functions $P_{ab}$ give the probability to find a parton $a$ inside a parton $b$. 
The splitting functions may be calculated from the Lagrangian of QCD.

Taking Mellin moments of both the parts of IDEs (\ref{SD}) and (\ref{GD}) we obtain matrix differential equation with the anomalous dimension matrix $\gamma_{ab}(N,\alpha),$  
where $N$ is a complex variable which corresponds to the Mellin moments. In the present article we take into account integrated gluon distribution $G(x,u)$ only. 
This approximation is known as a  dominant eigenvalue of the matrix of anomalous dimensions 
\cite{Ball:2005mj,Ball:1999sh,Altarelli:1999vw,Altarelli:2000dw,Altarelli:2000mh,Altarelli:2001ji,Altarelli:2003hk,Altarelli:2005ni} 
and may be justified in several gauge models.

The DGLAP IDE (\ref{differences}) may be written as a differential equation  for any pair $ij$ after taking the Mellin moments of both  the parts of it   
in  analogy  with  Eq. (\ref{RGEq}).  As it follows from (\ref{RGEq}), the anomalous dimension  $\gamma_{qq}(N,\alpha)$ is Mellin moment of the 
splitting functions $P_{qq},$ 
\begin{eqnarray}
M_{ij}[\Delta(x,u),x](N) = \int_0^1dx x^{N-1}\Delta_{ij}(x,u)  \no \\
u\frac{d}{du} M_{ij}[\Delta(x,u),x](N) = \frac{\alpha(u)}{2\pi}\int_0^1 dx x^{N-1}\int_x^1\frac{dy}{y}\Delta_{ij}(y,u)P_{qq}\left(\frac{x}{y}\right) \no \\
= \frac{\alpha(u)}{2\pi}\underbrace{\int_0^1 dx x^{N-1}P_{qq}(x)}_{\gamma_{qq}(N,\alpha)}\underbrace{\int_0^1 dy y^{N-1}\Delta_{ij}(y,u)}_{M_{ij}[\Delta(y,u),y](N)}. 
\end{eqnarray}

\subsection{About the model and DIS processes in this model} \label{sec:about}

Progress in the solution to DGLAP and BFKL equations has been achieved in ${\cal N}=4$ supersymmetric Yang-Mills theory 
\cite{Kotikov:2000pm,Kotikov:2002ab}. 
This is due to the fact that the gauge $\beta$-function vanishes in all loops in this theory. 
If supersymmetry in this model is softly broken, it would not spoil the vanishing of the  gauge $\beta$-function at the scale well above a gluino mass.  
However, the presence of a gaugino mass  may make superpartners heavy while the gluons remain massless 
\cite{Kazakov:1991th,Yamada:1994id,Kazakov:1995cy,Kondrashuk:1997uf,Jack:1997pa,Avdeev:1997vx,Kondrashuk:1999de,Kondrashuk:2000qb}
This would mean there is no running of the coupling in the model at the scale well above the threshold of gluino mass and due to this the confinement 
is not possible and the existence of nuclei is doubtful in ${\cal N}=4$ supersymmetric Yang-Mills theory. 
However, the bound states of three gluinos are possible in this model, they may serve as nuclei in the analysis of DIS processes 
in this field theory.  At the scale well below  the gluino mass threshold we have a pure QCD theory without fermions and with 
a running gauge coupling.

Massless gluons may be split into other massless gluons  
via the splitting functions $P_{GG},$ or in a superpartners via the splitting function   $P_{Gq}.$  The partonic model is described well 
by DGLAP equation. In this model there are three integrated parton distributions which are gluon distribution, gluino distribution and the corresponding 
scalar distribution. The corresponding solution to DGLAP IDE is a mixture of three power-like functions. However, there always is dominant 
contribution which has a dominant power. We treat integrated gluon distribution  $G(x,u)$ as this dominant contribution and do not take into account other two contributions
from gluino distribution and scalar distribution. This is a rough approximation to DGLAP IDE   of  ${\cal N}=4$ supersymmetric Yang-Mills theory.
However, it is a good model for searching analytical solution to this IDE. Such an analytical solution to DGLAP IDE is found in 
Section \ref{toy} and Section \ref{real} of the present article.  This approximation assumes that instead of matrix of anomalous dimensions we have only one 
function $\gamma(N,\alpha).$ Instead of DGLAP IDE  (\ref{GD}) for integrated gluon distribution we consider IDE (\ref{dglap approx}).

The evolution of integrated gluon distribution $G(x,u)$ is subject to the DGLAP IDE and the evolution of the unintegrated  gluon distribution is subject to the BFKL IDE. 
Both these IDEs must be consistent when applied to the gluon distribution which must satisfy them.   
The BFKL IDE is valid for each of three unintegrated distributions independently, however we consider it for unintegrated  gluon distribution only because in our model  we consider
DGLAP IDE only for  the unintegrated  gluon distribution.  We should consider DGLAP IDE and BFKL IDE in the kinematic region in which both 
equations are valid. We  show in the present article that DGLAP IDE is enough to find a general form of gluon distribution in the proposed toy-model and 
we do not need the BFKL IDE for this purpose. Also, the model described in the previous two paragraphs possesses a property that its gauge coupling does not run. 
There are many gauge theories that possess such a property \cite{Kazakov:1991th,Kazakov:1995cy,Jack:1997pa,Kondrashuk:1997uf,Jones:1986vp,Ermushev:1986cu,Avdeev:1992jt}.

\section{DGLAP equation with vanishing $\beta$-function for Integrated gluon distribution}

The integrated gluon distribution $G(x,u)$ is a dimensionless function, where $u=Q^2/\mu^2$, and $Q^2$ is the momentum transfer and $\mu^2$ a referential momentum transfer. 
It was constructed as one of the coefficient functions for the decomposition of  
the cross sections in DIS  processes in terms of the tensor structures. In the approximation described in the previous chapters  
this integrated gluon distribution satisfies the DGLAP IDE, that is,   
\begin{eqnarray} \label{dglap approx} 
u\frac{d}{du} G(x,u) = \frac{\alpha(u)}{2\pi}\int_x^1\frac{d y}{y}G(y,u)P_{GG}\left(\frac{x}{y},\alpha(u)\right), \\
u\frac{d}{du} G(N, u) = \frac{\alpha(u)}{2\pi}\gamma(N,\alpha(u))G(N,u), \no\\
G(N,u) = \int_0^{1}dx~x^{N-1}G(x, u), \no\\
\gamma(N,\alpha(u)) = \frac{\alpha(u)}{2\pi}\int_0^1 dx ~x^{N-1} P_{GG}\left(x,\alpha(u)\right). \no 
\end{eqnarray}

We take in this section $\alpha'(u) = 0$ and in the framework of this model we have the result for integrated gluon distribution 
\begin{eqnarray} \label{dglap G N u} 
u\frac{d}{d u}G\le N,u\ri = \frac{\alpha}{2\pi}\gamma(N,\alpha) G\le N,u\ri,   \Rightarrow   G\le N,u\ri =  G(N,1)u^{\dis{\frac{\alpha}{2\pi}\gamma(N,\alpha)}}, 
\end{eqnarray}
here $Q^2 = \mu^2$ ($u=1$) is a scale which corresponds to arbitrariness in solutions to differential equations. In Refs.\cite{Salam:1999cn,Kotikov:2002ab} it is 
called $Q_0^2$ scale.  We do not write any dependence on $\alpha$ in the integrated gluon distributions  $G\le N,u\ri,$ however in  $\gamma(N,\alpha)$ 
we write it explicitly. This will be useful for further expansions in terms of $\alpha.$

For the brevity, in the rest of the paper we will use the notation of Refs.\cite{Ball:1999sh}-\cite{Ball:2005mj}
$ G(N,u) \equiv M[G(x,u),x](N).$ From the theory of the 
integral transformations it follows  that the small $x$ region  for the dominant PDF $G(x,u)$ corresponds to the terms singular at the point $N=1$  of the Mellin moment  $M[G(x,u),x](N),$
see Section \ref{real}.

In this case we have a power-like dependence of PDFs on the momentum transfer \cite{Kotikov:2002ab}. Usually,
there are  some symmetry reasons to have the gauge coupling fixed. This happens for example in ${\cal N}=4$ supersymmetric Yang-Mills theory \cite{Brink:1976bc,Green:1982sw}, Chern-Simons non-Abelian topological Yang-Mills theory 
at fixed points of the renormalization group flows \cite{Avdeev:1992jt,Avdeev:1993ke}, finite supersymmetric Yang-Mills theories with low level of supersymmetry \cite{Ermushev:1986cu,Jones:1986vp},
softly broken finite Yang-Mills theories \cite{Kazakov:1991th,Yamada:1994id,Kazakov:1995cy,Kondrashuk:1997uf,Avdeev:1997vx,Kondrashuk:1999de,Kondrashuk:2000qb}.  In  ${\cal N}=4$ supersymmetric Yang-Mills theory 
twist-two operators may be combined in representations irreducible with respect to the renormalization group with the property of multiplicative renormalization \cite{Kotikov:2002ab}, 
and even in supersymmetric theories with the lower level of supersymmetry  a dominant PDF may exist in the small $x$ limit \cite{Almasy:2011eq,Kotikov:2020ukv}. 
We may expect that that, if an irreducible with respect to the renormalization group multiplicatively renormalizable combination of Mellin moments of PDFs contains the moment of gluon PDF,
than it is  dominant in the small $x$ limit represented by the terms singular at the point $N = 1$ of the complex plane of the Mellin variable.  A number of these involved irreducible representations of the Mellin moments of PDFs  
which are dominant in the region $N \rightarrow 1$ depends on the level of symmetry of the theory in this limit for a given theory\footnote{Normalization of the PDFs in Refs. 
\cite{Ball:1999sh,Altarelli:1999vw,Altarelli:2000dw,Altarelli:2000mh,Altarelli:2001ji,Altarelli:2003hk,Altarelli:2005ni,Ball:2005mj} is different and the singularity of the gluon PDF at small $x$ 
corresponds to the point $N=0$ in the complex plane of the Mellin variable.}.

The solution of the DGLAP equation for the running coupling for the integrated PDF is given in Appendix \ref{Run} and it  has been partially considered in Ref. \cite{Kondrashuk:2019cwi}
This paper is mainly dedicated to the fixed coupling so that all the comments on the case of the running coupling were put in Appendices.

\section{DGLAP equation with vanishing $\beta$-function for Unintegrated gluon distribution} \label{Updf}

It is known that integrated gluon distribution $G(x,u)$  is related to unintegrated  gluon distribution  $\varphi(x,k_\perp^2)$ via the integral relation 
\begin{eqnarray} \label{UGD}
G\le x,\frac{Q^2}{\mu^2}\ri = \int_0^{Q^2}dk_\perp^2 \varphi\le x,\frac{k_\perp^2}{\mu^2}\ri, 
\end{eqnarray}
here  $\varphi\le x,{k_\perp^2}/{\mu^2}\ri$ is the unintegrated  dominant PDF. 
It appears that it is always possible to construct from $\varphi\le x,{k_\perp^2}/{\mu^2}\ri$ a function which satisfies the same DGLAP equation  (\ref{dglap approx}) as well as the integrated  $G(x,u)$ dominant PDF does.
In Section \ref{Updf} and  in Appendix \ref{Updr} we show this statement is true in both the cases of the fixed  (Section \ref{Updf}) and of the running gauge coupling (Appendix  \ref{Updr}).  
We need to consider the unintegrated PDF because the dual IDE which is called the BFKL equation is written for unintegrated PDFs \cite{Lipatov:1976zz,Fadin:1975cb,Kuraev:1976ge,Kuraev:1977fs,Balitsky:1978ic}.
We may get this dual DGLAP equation (BFKL equation) via a complex diffeomorphism from the DGLAP equation, as it has been done in Ref. \cite{Kondrashuk:2019cwi}. 
This means these two IDEs, DGLAP and BFKL, should be written for the same quantities that are the unintegrated PDFs.

From Eq.(\ref{UGD}) we conclude that their Mellin moments are related too by the same integral relation  
\begin{eqnarray}
G\le N,\frac{Q^2}{\mu^2}\ri = \int_0^{Q^2}dk_\perp^2 \varphi\le N,\frac{k_\perp^2}{\mu^2}\ri, 
\end{eqnarray}
where we denoted
\begin{eqnarray}
\varphi \le N,\frac{k_\perp^2}{\mu^2} \ri = \int_0^{1}dx~x^{N-1} \varphi\le x, \frac{ k_\perp^2}{\mu^2}\ri.  
\end{eqnarray}
In turn, this unintegrated  gluon distribution  $\varphi(x,k_\perp^2)$ solves the BFKL equation.   
In  maximally supersymmetric Yang-Mills theory together with this function other unintegrated distributions 
like fermionic gluino distribution and scalar distribution exist \cite{Kotikov:2002ab}.  
Integrated gluon distribution is dimensionless function and unintegrated  gluon distribution is dimensionful function.

From Eq.(\ref{dglap G N u}) we obtain 
\begin{eqnarray} \label{UGD-2}
Q^2\frac{d}{dQ^2} \int_0^{Q^2}dk_\perp^2 \varphi\le N,\frac{k_\perp^2}{\mu^2}\ri = \frac{\alpha}{2\pi}
\gamma(N,\alpha) \int_0^{Q^2}dk_\perp^2 \varphi\le N,\frac{k_\perp^2}{\mu^2}\ri ,  \Rightarrow   \no \\
Q^2\frac{d}{dQ^2} Q^2\varphi\le N,\frac{Q^2}{\mu^2}\ri = \frac{\alpha}{2\pi}\gamma(N,\alpha)   Q^2  \varphi\le N,\frac{Q^2}{\mu^2}\ri.
\end{eqnarray}
This simple transformation shows that the dimensionless function $\dis{Q^2\varphi\le N, Q^2/\mu^2\ri}$ satisfies 
the same DGLAP equation as  Mellin moments $\dis{G \le N, {Q^2}/{\mu^2}\ri} $ of integrated gluon distribution $\dis{G \le x, {Q^2}/{\mu^2}\ri} $   
do, and with the same power-like solution 
\begin{eqnarray} 
Q^2\varphi \le N,\frac{Q^2}{\mu^2}\ri =  \mu^2 \varphi(N,1) \le\frac{Q^2}{\mu^2}\ri^{\dis{\frac{\alpha}{2\pi}\gamma(N,\alpha)}}.  
\end{eqnarray}
A new function  $\phi \le N,\dis{\frac{Q^2}{\mu^2}} \ri$  may be introduced for the future use 
\begin{eqnarray} \label{phi N u gamma}
Q^2 \varphi \le N, \frac{Q^2}{\mu^2}\ri \equiv \phi \le N,\frac{Q^2}{\mu^2} \ri =  \phi(N,u) = \phi(N,1) u^{\dis{\frac{\alpha}{2\pi}\gamma(N,\alpha)}} 
\equiv  \phi_1(N) \le \frac{Q^2}{\mu^2} \ri^{\dis{\frac{\alpha}{2\pi}\gamma(N,\alpha)}}.
\end{eqnarray}
This new function $\phi \le N,u\ri$ is Mellin $N$-moment of the solution to the DGLAP IDE 
\begin{eqnarray} 
u \frac{d}{d u} \phi \le x,u\ri = \frac{\alpha}{2\pi}\int_x^1\frac{d y}{y}\phi \le y, u\ri P_{GG}\le \frac{x}{y},\alpha\ri,    \label{dglap phi classic} \\
u \frac{d}{d u} \phi \le N, u\ri   = \frac{\alpha}{2\pi}\gamma(N,\alpha) \phi\le N, u\ri, \label{dglap phi N u meq} \\
\phi \le N,u\ri = \int_0^{1}dx~x^{N-1}\phi\le x, u\ri,  \label{dglap phi N phi x u} \\
\gamma(N,\alpha) = \int_0^1 dx ~x^{N-1} P_{GG}\left(x,\alpha \right), \label{dglap gamma}
\end{eqnarray}
and for this IDE the domain of $u$ is a real nonnegative $u \in [0, \infty[.$ In order to uniform notation with Appendix \ref{Updr} dedicated to the running coupling 
we change the normalization of the dimensionless function  $\dis{\phi\le N,{Q^2}/{\mu^2}\ri}$  by a factor which is a simple constant when the coupling does not run, 
\begin{eqnarray*} 
\phi\le N,\frac{Q^2}{\mu^2}\ri = Q^2\varphi\le N, \frac{Q^2}{\mu^2}\ri  \longrightarrow \phi\le N,\frac{Q^2}{\mu^2}\ri = \frac{2\pi~~  Q^2 \varphi\le N, Q^2/\mu^2 \ri}{\alpha\gamma(N,\alpha) }.
\end{eqnarray*}
After this renormalization, we may show 
\begin{eqnarray*} 
\phi\le N,1\ri \equiv   \phi_1\le N \ri =  G(N,1),
\end{eqnarray*}
that is, the shape function $\phi\le N,1\ri$ of the unintegrated dominant PDF is parametrized the same way as the  shape function  $G(N,1)$ of its integrated dominant PDF is.

It may be shown that a self-consistency condition should be imposed on the shape function $\phi_1\le N \ri$ which may be obtained directly from the DGLAP equation in its integro-differential form. In Section \ref{sec:method to solve}
it is shown that such self-consistency conditions may be written for the frozen and for the running coupling. These conditions almost coincide for the cases of the running and of the fixed coupling.    
The self-consistency condition for the shape function in the case of the frozen coupling is applied in Sections \ref{toy} and \ref{real}.   
The self-consistency condition in the case of the running coupling has been obtained in Appendix \ref{Sccr} by completely the same method as it has been done in the case of the fixed coupling.

\section{Contour of the inverse transformation from $N$ to $x$ }

The domain of variable $x$ of  $\phi \le x,u\ri$ should include the interval $x \in [0, 1],$ otherwise the transformation to Mellin moment   (\ref{dglap phi N phi x u})
would be impossible to define. In brief, summarizing the discussion of the previous section, 
if we know  (\ref{dglap phi N phi x u})  then to recover $\phi \le x, u \ri$ when $x \in [0, 1]$
we need to make the inverse transformation (\ref{MMT Cauchy}) via Cauchy formula\footnote{Here we should mention that any inverse integral transformation 
obtained by Cauchy formula in our paper includes factor ${1}/{2\pi i}.$ 
We do not write it for the brevity.} 
\begin{eqnarray} \label{MM phi C1} 
\phi(x,u) =  \int_{a+\delta-i\infty}^{a+\delta+i\infty} ~dN  x^{-N} \phi(N,u) 
\end{eqnarray}

It is supposed that Mellin moment $\phi(N,u)$ is defined in the domain  ${\rm Re}~N > a,$ where $a$ is an index of the power-like growth of 
the function $\phi(x,u),$ 
\begin{eqnarray}
\phi(x,u) < A/x^{a}. 
\end{eqnarray}
In the inverse transformation (\ref{MM phi C1})  the contour passes vertically in the complex plane $N$ at ${\rm Re}~N = a + \delta.$ 
Under this condition the  $\phi(N,u)$ does not have poles in the complex half-plane to the right from this vertical line 
in the complex plane of variable $N.$

\section{Method to solve the DGLAP equation analytically} \label{sec:method to solve}

In this Section we propose how DGLAP IDE may be solved without making use of the BFKL equation. This may be considered as an alternative way to the 
approach of Refs.\cite{Ball:1999sh}-\cite{Ball:2005mj}  and to the approach of Refs.\cite{Kotikov:2000pm,Kotikov:2002ab}.  As we have mentioned in Introduction, the use of BFKL 
was a trick there to get some information about possible solution to DGLAP equation. One of the motivations for these approaches was that BFKL kernel is better 
known than DGLAP kernel and it was more easy to calculate the BFKL kernel than to calculate the DGLAP kernel \cite{Kotikov:2000pm,Kotikov:2002ab,Fadin:1998py} 
at the same loop order. 

DGLAP IDE (\ref{dglap phi classic}) has a solution in the form  of Eq. (\ref{phi N u gamma}) 
for Mellin $N$-moment $\phi(N,u).$ This solution does not restrict the form of function $\phi_1(N).$  The reason is that when we do the 
integration over variable $x$ on both sides of IDE (\ref{DGLAP-1}), we are averaging the information about $x$ in unintegrated  gluon distribution $\phi(x,u).$
After this averaging we obtain differential equation for the Mellin moments like Eqs. (\ref{RGEq}), (\ref{dglap G N u}) and (\ref{dglap phi N u meq}).

However, we may look at DGLAP IDE at a different angle and substitute the inverse transformation (\ref{MM phi C1}) in DGLAP 
IDE (\ref{dglap phi classic}) for unintegrated  gluon distribution  $\phi(x,u).$ Such a strategy  should give restrictions on function  $\phi_1(N),$ because we use pointwise information.   
Indeed, by doing this we obtain 
\begin{eqnarray} \label{conclusion-11} 
u\frac{d}{d u} \phi \le x,u\ri = \frac{\alpha}{2\pi}\int_x^1\frac{d y}{y}
\phi \le y, u\ri P_{GG}\le \frac{x}{y},\alpha\ri   \Rightarrow  \no\\
u\frac{d}{d u}\int_{a-i\infty}^{a+i\infty}~dN x^{-N} \phi(N,u) = \frac{\alpha}{2\pi}\int_x^1\frac{d y}{y}\int_{a-i\infty}^{a+i\infty}~dN y^{-N} \phi(N,u) 
P_{GG}\le \frac{x}{y},\alpha\ri  \Rightarrow \no\\
\int_{a-i\infty}^{a+i\infty}~dN x^{-N} \phi_1(N)u^{\dis{\frac{\alpha}{2\pi}\gamma(N,\alpha)}}\gamma(N,\alpha) = \no\\
= \int_x^1\frac{d y}{y}
\int_{a-i\infty}^{a+i\infty}~dN y^{-N}  \phi_1(N)u^{\dis{\frac{\alpha}{2\pi}\gamma(N,\alpha)}}
P_{GG}\le \frac{x}{y},\alpha\ri  \Rightarrow \no\\
\int_{a-i\infty}^{a+i\infty}~dN x^{-N} \phi_1(N)u^{\dis{\frac{\alpha}{2\pi}\gamma(N,\alpha)}}\left[\gamma(N,\alpha) - x^N
\int_x^1\frac{d y}{y} y^{-N}  P_{GG}\le \frac{x}{y},\alpha\ri \right] = 0
\end{eqnarray} 
The integral in the bracket may be transformed to    
\begin{eqnarray} \label{conclusion-21} 
\int_x^1 \frac{dy}{y} ~y^{-N} P_{GG}\le \frac{x}{y},\alpha\ri = \int_1^{1/x} \frac{dy}{y} ~y^{N} P_{GG}(xy,\alpha) 
= x^{-N}\int_x^1 \frac{dy}{y} y^{N} P_{GG}(y,\alpha).
\end{eqnarray}
The DGLAP IDE may be written in such a form  
\begin{eqnarray} \label{dglap T2} 
\int_{a-i\infty}^{a+i\infty}~dN x^{-N} \phi_1(N)u^{\dis{\frac{\alpha}{2\pi}\gamma(N,\alpha)}}\left[\gamma(N,\alpha) - \int_x^1 \frac{dy}{y} y^{N} P_{GG}(y,\alpha)\right] \no \\
= \int_{a-i\infty}^{a+i\infty}~dN x^{-N} \phi_1(N)u^{\dis{\frac{\alpha}{2\pi}\gamma(N,\alpha)}}\int_0^x \frac{dy}{y} y^{N} P_{GG}(y,\alpha) = 0
\end{eqnarray}
For the future use we introduce the notation
\begin{eqnarray} \label{T} 
T(N,x,\alpha) \equiv  \int_x^1 \frac{dy}{y} y^{N} P_{GG}(y,\alpha).  
\end{eqnarray}

The main idea is the contour integral should be put to zero in front of each power of expansion in terms of $x$ on the right hand side of Eq. (\ref{dglap T2})  for the same contour.

The method we have proposed in this Section is based on the fact 
that integrals of the splitting functions in the range from $0$ till $x$ (where $x$ is  Bjorken variable) are proportional to $x^N$ where $N$ is the complex variable of the 
Mellin moment $\phi(N,Q^2/\mu^2)$ of the unintegrated  dominant PDF $\phi(x,Q^2/\mu^2).$
Due to cancellation of this power $x^N$ with the power $x^{-N}$ which stands in the inverse integral transformation, 
we obtain an expansion in terms of integer powers of $x$ from which we may conclude that the coefficient in front of each integer power of $x$ must be zero. These requirements give us 
a set of integrals involving the Mellin moment $\phi(N,Q^2/\mu^2)$  of the unintegrated  dominant PDF  $\phi(x,Q^2/\mu^2)$  which must be equal to zero simultaneously. 
In the next Sections  \ref{toy} and \ref{real}  we have substituted the 
inverse Mellin moment  $\int_{a-i\infty}^{a+i\infty}~dN x^{-N} \phi(N,u)$   into this DGLAP equation (\ref{dglap phi classic}) and have obtained the equation (\ref{conclusion-11}) for the case of the frozen coupling,
which may be treated as a self-consistency condition for the shape of the PDF.  In Appendix \ref{Sccr} we simply repeat this trick  for the case of the running coupling.

\section{Solution to DGLAP equation in a simple toy-model} \label{toy}

The IDEs of the type like Eq. (\ref{DGLAP-1}) or in particular Eq. (\ref{dglap phi classic}) have a probabilistic interpretation and appear in many areas of applied 
mathematics, mathematical biology, or stochastic processes in theoretical chemistry \cite{chemestry}. Some of the authors of DGLAP IDE mentioned on page 321 of textbook \cite{Ioffe:2010zz} 
that this equation is analogous to balance equation of various gases being in chemical equilibrium. It is not necessary that there exists a quantum field theory model for any given splitting function $P(z).$
Quantum field theory is not the unique field  of application for this IDE. 
The existence of a wide spectrum of applications suggests that analytical solution to such a type of IDEs should be searched. The splitting function $P(z)$ is an input for 
this IDEs. In this Section we take the splitting function in the simplest form of only one term in order to show that the method we have found works for solving this IDE. 
Almost realistic form of the splitting function $P_{GG}(z)$ will be considered in the next Section.

We consider in this Section the splitting function of gluons in the form 
\begin{eqnarray} \label{toy model} 
P_{GG}(z,\alpha) =  \beta_0\delta(1-z) + 2z
\end{eqnarray}
With this simple splitting function we may illustrate the main idea of the method. First, according to Eq. (\ref{T}) we have  
\begin{eqnarray} 
T(N,x,\alpha) =  \int_x^1 \frac{dy}{y} y^{N} P_{GG}(y,\alpha) = \int_x^1 \frac{dy}{y} y^{N} (\beta_0\delta(1-y) + 2y) = 
  \beta_0 + \frac{2}{N+1} -  \frac{2x^{N+1}}{N+1}.
\end{eqnarray}
We have from Eqs. (\ref{dglap gamma}) and (\ref{T})  
\begin{eqnarray}\label{gamma toy} 
\gamma(N,\alpha) = T(N,0,\alpha) =  \beta_0 + \frac{2}{N+1}.
\end{eqnarray}
Thus, Eq. (\ref{dglap T2}) may be rewritten in this case as 
\begin{eqnarray} 
\int_{a-i\infty}^{a+i\infty}~dN x^{-N} \phi_1(N)u^{\dis{\frac{\alpha}{2\pi}\gamma(N,\alpha)}}  \frac{x^{N+1}}{N+1}  =  x\int_{a-i\infty}^{a+i\infty} 
~dN \frac{\phi_1(N)u^{\dis{\frac{\alpha}{2\pi}\gamma(N,\alpha)}}}{N+1} = 0, 
\end{eqnarray}
from which we must conclude 
\begin{eqnarray} \label{dglap x degree} 
\int_{a-i\infty}^{a+i\infty}~dN \frac{\phi_1(N)u^{\dis{\frac{\alpha}{2\pi}\gamma(N,\alpha)}}   }{N+1} = 0. 
\end{eqnarray}

Eq. (\ref{dglap x degree}) does not restrict unintegrated  gluon distribution  $\phi(x,u)$ completely. Indeed, as we have explained in the previous sections, our model 
suggests that gluon distribution is the dominant distribution in ${\cal N} = 4$ supersymmetric Yang-Mills theory. This is a rough approximation under which we suppose that the 
gauge coupling does not run and gaugino and scalar distribution are not taken into account. Coefficient $\beta_0$ is the first coefficient  
of the gauge $\beta$ function. Since $\beta=0$ (the coupling does not run), we have $\beta_0=0$. Then, Eq.  (\ref{dglap x degree})  takes the form 
\begin{eqnarray} 
0 = \int_{a-i\infty}^{a+i\infty}~dN \frac{\phi_1(N) u^{\dis{\frac{\alpha}{2\pi}\gamma(N,\alpha)}} }{N+1} =   \int_{a-i\infty}^{a+i\infty}~dN \frac{\phi_1(N)u^{ \dis{\frac{\alpha}{\pi}\frac{1}{N+1}  }} }{N+1} = \no\\
\frac{\pi}{\alpha}{u\partial_u} \int_{a-i\infty}^{a+i\infty}~dN \phi_1(N)u^{ \dis{\frac{\alpha}{\pi}\frac{1}{N+1}  }}  = 
\frac{\pi}{\alpha}\le{u\partial_u} \int_{a-i\infty}^{a+i\infty}~dN \phi_1(N)u^{ \dis{\frac{\alpha}{\pi}\frac{1}{N+1}  }}x^{-N}\ri_{x=1} = \no\\
\frac{\pi}{\alpha}\le{u\partial_u} \phi(x,u)\ri_{x=1} = \frac{\pi}{\alpha}{u\partial_u} \phi(1,u). 
\end{eqnarray}
There are many functions satisfying this condition. For example, any expansion in powers of $\ln x$
\begin{eqnarray} 
\phi(x,u) = C + \sum_{k=1}^{\infty}f_k(x,u)\ln^k{x}  
\end{eqnarray}
where $f_k(x,u)$ are non-singular functions of $x$ at $x=1,$ would work as gluon distribution satisfying Eq. (\ref{dglap x degree}).

We use Eq. (\ref{dglap x degree}) to fix point $a$ on the real axis in the complex plane of variable $N$ and to find function $\phi_1(N).$  
First, we go back to Eq. (\ref{dglap x degree}) and expand it in power of $\ln{u}.$ This expansion helps to establish the value of $a,$ indeed, 
\begin{eqnarray} \label{dglap-4} 
0 = \int_{a-i\infty}^{a+i\infty}~dN \frac{\phi_1(N)u^{\dis{\frac{\alpha}{2\pi}\gamma(N,\alpha)}} }{N+1} =   
\int_{a-i\infty}^{a+i\infty}~dN \frac{\phi_1(N)u^{ \dis{\frac{\alpha}{\pi}\frac{1}{N+1}  }} }{N+1} =  \no\\
\sum_{k=0}^{\infty} \frac{1}{k!}\le\frac{\alpha}{\pi}\ln{u}\ri^k \int_{a-i\infty}^{a+i\infty}~dN \frac{\phi_1(N)}{(N+1)^{k+1}} \Rightarrow \no\\  
\int_{a - i \infty}^{a + i \infty} d N \frac{\phi_1(N)}{(N+1)^{k+1}} = 0 \qquad (k=0, 1, 2, \ldots) 
\end{eqnarray}
According to the theory of transformation to Mellin moment  described in Section 2, all the poles should be situated to the left from the point $N =a$ in the complex plane of variable $N$ ($\Rightarrow -1 < a$), 
and the contour should be closed to the negative complex infinity because $x \in [0,1].$  There are two different possibilities to guarantee zero on the r.h.s. of Eq. (\ref{dglap-4}). The first possibility is that all 
the poles should be of second order or higher in order to avoid contribution of residues  due to Cauchy formula. This means that all the poles should be at the same point. In this Section dedicated to 
a simple toy-model we concentrate on this first possibility. Another possibility when residues at two different points cancel each other 
is considered in the next Section in which we study the solution to DGLAP by this method for almost realistic splitting function $P_{GG}(z).$

We have already the pole at the point $N= -1$  in Eq. (\ref{dglap x degree}). Going along the first way described in the previous paragraph in order to solve Eq.~(\ref{dglap-4}) we choose that 
\begin{eqnarray} 
\phi_1(N)  =  \sum_{j=1}^{\infty}\frac{c_j}{(N+1)^{j}}, 
\end{eqnarray}
where $c_j$ are arbitrary coefficients.  Another conclusion is that $a$ is situated to the right from $N=-1$ on the real axis because $x \in [0,1]$ and the contour should be closed to the left,
that is, $-1 < a.$  In such a case the poles at the point $N=-1$ will be taken into account when we use Cauchy integral formula to calculate unintegrated gluon distribution  $\phi(x,u).$
We conclude that Eq. (\ref{dglap x degree}) is enough to fix the contour and contains good piece of information about function $\phi_1(N).$ 
The function $\phi_1(N)$ could have, in fact, also terms of the form $\prod_{j=1}^{n} (N-N_j)^{-\nu_j}$ with ${\rm Re}(N_k) \leq -1$ 
and at least one natural power index being $\nu_k$ positive nonzero, as argued in a more general context in detail in the next Section.

As an example, we may obtain the form of unintegrated  gluon distribution $\phi(x,u)$ for the simplest case when $\phi_1(N)  =  1/(N+1),$ that is, 
\begin{eqnarray} \label{Examp}
\phi(x,u) = \int_{-1+\delta-i\infty}^{-1+\delta+i\infty}~dN \phi_1(N)u^{ \dis{\frac{\alpha}{\pi}\frac{1}{N+1}  }}x^{-N} = 
\int_{-1+\delta-i\infty}^{-1+\delta+i\infty}~dN \frac{x^{-N}}{N+1} u^{ \dis{ \frac{\alpha}{\pi} \frac{1}{N+1} }} = \no \\
\sum_{k=0}^{\infty}\frac{1}{k!}\le\frac{\alpha}{\pi}\ln{u}\ri^k \int_{-1+\delta-i\infty}^{-1+\delta+i\infty}~dN \frac{x^{-N}}{(N+1)^{k+1}} =  \no \\
x\sum_{k=0}^{\infty}\frac{1}{k!}\le\frac{\alpha}{\pi}\ln{u}\ri^k \int_{-1+\delta-i\infty}^{-1+\delta+i\infty}~dN \frac{x^{-N-1}}{(N+1)^{k+1}} \no \\
= x\sum_{k=0}^{\infty}\frac{1}{k!}\le\frac{\alpha}{\pi}\ln{u}\ri^k \frac{(-\ln{x})^{k}}{k!} =  x\sum_{k=0}^{\infty}\frac{(-1)^{k}}{(k!)^2}
\le\frac{\alpha}{\pi}\ln{u}\ln{x}\ri^k = xI_0\le 2\sqrt{\frac{\alpha}{\pi}\ln{u}\ln{\frac{1}{x}}} \ri,
\end{eqnarray}
where $I_0$ is the modified Bessel function. On this side we reproduce Bessel-like behaviour obtained in Ref.\cite{Salam:1999cn} by summation of ladder diagrams
in the pure gluonic case too. However, the Bessel-like behaviour has been obtained in Ref.\cite{Salam:1999cn} under some approximations for the realistic gluon splitting function
$P_{GG}$ of Eq.(\ref{almost realistic model}). Our toy-model gives an exact solution for the Bessel-like behaviour with the one-term splitting function  (\ref{toy model}).  

To check that the function we found possesses necessary upper bounds on its behaviour with respect to variable $x,$ we do a simple approximation  
\begin{eqnarray} 
x\sum_{k=0}^{\infty}\frac{(-1)^{k}}{(k!)^2}\le\frac{\alpha}{\pi}\ln{u}\ln{x}\ri^k  \leqslant x\sum_{k=0}^{\infty}\frac{(-1)^{k}}{k!}\le\frac{\alpha}{\pi}\ln{u}\ln{x}\ri^k = 
xe^{\dis{{-\frac{\alpha}{\pi}\ln{u}\ln{x}}}} = x^{\dis{1-\frac{\alpha}{\pi}\ln{u}  }}
\end{eqnarray}

In arbitrary case we obtain
\begin{eqnarray} \label{general case} 
\phi(x,u) = \int_{-1+\delta-i\infty}^{-1+\delta+i\infty}~dN \phi_1(N)u^{ \dis{\frac{\alpha}{\pi}\frac{1}{N+1}  }}x^{-N} = \sum_{j=1}^{\infty}c_j
\int_{-1+\delta-i\infty}^{-1+\delta+i\infty}~dN \frac{x^{-N}}{(N+1)^j} u^{ \dis{ \frac{\alpha}{\pi} \frac{1}{N+1} }}  \no\\
= \sum_{j=1}^{\infty}\sum_{k=0}^{\infty}\frac{c_j}{k!}\le\frac{\alpha}{\pi}\ln{u}\ri^k \int_{-1+\delta-i\infty}^{-1+\delta+i\infty}~dN \frac{x^{-N}}{(N+1)^{j+k}} \no \\
= x  \sum_{j=1}^{\infty}\sum_{k=0}^{\infty}\frac{c_j}{k!}\le\frac{\alpha}{\pi}\ln{u}\ri^k \int_{-1+\delta-i\infty}^{-1+\delta+i\infty}~dN \frac{x^{-N-1}}{(N+1)^{j+k}} \no\\
= x \sum_{j=1}^{\infty} \sum_{k=0}^{\infty}\frac{c_j}{k!}\le\frac{\alpha}{\pi}\ln{u}\ri^k \frac{(-\ln{x})^{k+j-1}}{(k+j-1)!} =  
x\sum_{j=1}^{\infty} \sum_{k=0}^{\infty}\frac{(-1)^{k+j-1}c_j}{k!(k+j-1)!}\le\frac{\alpha}{\pi}\ln{u}\ln{x}\ri^k \ln^{j-1}{x} = \no\\
x\sum_{j=1}^{\infty} \sum_{k=0}^{\infty}\frac{c_j}{k!(k+j-1)!}\le\frac{\alpha}{\pi}\ln{u}\ln{\frac{1}{x}}\ri^k \ln^{j-1}{\frac{1}{x}}
\end{eqnarray}
This is the general solution to DGLAP IDE (\ref{dglap phi classic}) with the splitting function (\ref{toy model}). As we may observe, the solution is not unique. 
There are infinitely  many constants $c_j$ which appear in this solution.

Such toy-models remain to be useful practically even nowadays  because may capture in a compact expression the behaviour of a given asymptotic regime in QCD, In particular, the model (\ref{Examp})
possesses the Bessel-like behaviour with respect to square root of the product of logarithm on the Bjorken variable and logarithm of the momentum transfer  in the region of the small values of  $x$ 
when the main contribution comes from the gluon part  of the matrix DGLAP equation. Although the computational progress of the last decades is impressive  
(see for example Refs. \cite{Moch:2004pa,Vogt:2004mw,Botje:2010ay,Botje:2016wbq}) and  the perturbative solution to the DGLAP equation is already 
computed up to N$^2$LO for the Mellin moments of parton distribution functions with full inclusion of running coupling, 
the approximate solutions to the DGLAP equation corresponding to simple models still may help a lot in order to estimate physical quantities in the limits in which  numerical 
tools and solutions show bad behaviour in the practical models like QCD.

In addition to serve as a consistency check for the numerical or analytical 
calculation based on a powerful software, the approximate solutions to DGLAP IDE which are presented by the models considered in this Section may be used to train neural networks \cite{Alvarez:2019eaa}.      
Indeed, global analysis of the parton distribution functions taking into account recent data from the LHC is made by several scientific groups in the world \cite{Hou:2019efy,Dulat:2015mca,nnpdf-1,nnpdf-2,Ethier:2020way}. 
Many PDF parameters of initial parton distribution functions may be fixed from data only because  they cannot be computed from first principles. 
The software for the fitting of the  PDF parameters and for the PDF evolution is created  on the principles of neural networks \cite{nnpdf-1,nnpdf-2} which are an efficient tool 
to treat a big amount of data. The forms of parton distribution functions at some scale used in such a fitting procedure tend to be some combination of Euler beta functions
\cite{Ball:2016spl,Alekhin:2003ev,Blumlein:2006ws,Alekhin:2012ig} which than evolve from that scale according to the DGLAP integro-differential equation. 

Also, these models may be used for developing  alternative analytical methods to calculate the contour integrals which appear in the inverse Mellin transformation. In Ref. \cite{Alvarez:2019eaa} 
such contour integrals have been transformed via diffeomorphism in the complex plane of the Mellin moment variable to the contour integrals of the inverse Laplace transformation 
of the Jacobian of the corresponding complex map. In turn, these contour integrals of the inverse Laplace transformation may be represented in terms of the Barnes integrals 
by deforming the Hankel contour in the complex plane \cite{Alvarez:2019eaa}.

\section{Solution to DGLAP IDE in almost realistic case} \label{real}

In the previous Section a toy-model has been considered. The idea was to show how the method proposed in Section \ref{sec:method to solve} works. The method was aimed to 
solve integro-differential equations of the DGLAP type, like Eq. (\ref{DGLAP-1}) or in particular Eq. (\ref{dglap phi classic}). These equations have a probabilistic interpretation
and due to this interpretation have many practical applications in science and technology.

The toy-model was chosen to be simple, it contains one term only. However, for this toy-model we have reproduced Bessel-like behaviour of the unintegrated gluon distribution of  
Ref.\cite{Salam:1999cn} in which such a kind of behaviour has been obtained via an estimative summation of the ladder diagrams in pure gluonic QCD with the gluon splitting 
function  $P_{GG}$ given in Eq.(\ref{almost realistic model}). 
This gluon splitting function $P_{GG}(x)$ of  Eq.(\ref{almost realistic model}) has been calculated  at the one-loop level and may be found in many textbooks.

In contrast to Ref.\cite{Salam:1999cn},  we take the $\delta$-function term in this splitting function equal to zero. This is because the coupling in our model does not run.
This model comes from maximally supersymmetric Yang-Mills theory in which supersymmetry is softly broken. The model  is described in   Section \ref{sec:about}. 
In this model we take the contribution of gluon distribution only on the r.h.s. of the DGLAP IDEs and neglect the contribution of gluino and scalar distributions.
This is a rough approximation under which we suppose that the gauge coupling does not run and at the same time gaugino and scalar distribution are not taken into account.
The unintegrated gluon distribution looks to be the dominant distribution in this model. This would be almost realistic model. 
Knowing solution in this case, we may get an impression how the gluon distribution looks in a realistic model in which all three distribution would participate.

The explicit form of the realistic  gluon splitting function $P_{GG}(z)$ may be found in any texbook dedicated to QCD or to Quantum Field Theory in general (for example in Ref. \cite{Quigg:2013ufa}, page 236, Eq. (8.5.42)), 
or in the original paper \cite{Altarelli:1977zs}, 
and it takes the form  
\begin{eqnarray} \label{almost realistic model} 
P_{GG}(z) = 2C_2(G)\left[\frac{z}{(1-z)_{+}}  + \frac{1-z}{z}  + z(1-z) + \beta_0\delta(1-z) \right],
\end{eqnarray}
in which $\beta_0$ is the one-loop coefficient of the gauge $\beta$-function.

We have to put $\beta_0=0$ because the coupling does not run in the case that we consider in this paper. This point requires a special comment. In ${\cal N}=4$ supersymmetric Yang-Mills theory 
the coupling does not run to all the loops. However, Eq. (\ref{almost realistic model}) is just a leading-order contribution to the splitting function $P_{GG}(z).$ 
In the original papers of \cite{Altarelli:1977zs,Dokshitzer:1977sg} the splitting function $P_{GG}(z)$   corresponds to the kernel of Bethe-Salpeter equation \cite{Dokshitzer:1977sg}.  
We do not consider higher-order corrections to the splitting function  $P_{GG}(z)$ in the present paper. Thus, the solution to the DGLAP equation with the splitting function (\ref{almost realistic model})
is the solution but only at the leading order. Its order is determined by the order of the splitting function.  We do not consider other splitting functions due to the reasons that we have explained in the previous Sections.
The gluon distribution dominates in the small $x$ limit in QCD and in the conformal gauge theory like ${\cal N}=4$ supersymmetric Yang-Mills theory.

Altarelli and Parisi  in Ref. \cite{Altarelli:1977zs} have shown that the approach based on the operator product expansion used in the Nobel prize paper  \cite{Gross:1974cs} admits a probabilistic interpretation 
in terms of the splitting functions (\ref{almost realistic model}).  It was found in Ref.\cite{Altarelli:1977zs}  that these splitting functions are consistent with 
the anomalous dimensions  of the twist two operators calculated in \cite{Gross:1974cs}. Similar splitting functions appeared in the approach of  
Refs. \cite{Gribov:1972ri,Gribov:1972rt,Dokshitzer:1977sg} based on the Bethe-Salpeter equation imposed on the contributing family of Feynman diagrams.

The coefficient $2C_2(G)$ in the expression for the splitting function (\ref{almost realistic model}) is actually $2N$ for the gauge group $SU(N)$ \cite{Altarelli:1977zs}. 
For QCD, for example, we consider the group $SU(3).$ 
Thus it is a universal coefficient based on the gauge group contribution, it does not depend on the representation of the quark fields. However, the coefficient  $\beta_0$ is very sensitive to the 
representation of the matter fields. In QCD this coefficient is responsible for the phenomenon of the asymptotic freedom \cite{Gross:1974cs}.

The solution to the DGLAP IDE for the Mellin moment of the dominant parton distribution is given 
in Appendices \ref{Run} and \ref{Updr}. At the leading order of the perturbation theory   for the case of the running coupling the solution 
to the  DGLAP IDE may be represented in the  same form of the contour integral (\ref{Examp}) which we obtained for the case of the fixed coupling. 
The  only difference with the fixed coupling case is that instead of the power function of $u$ in the integrand of (\ref{Examp})  another dependence on the momentum transfer $u$ will stand.
At higher orders of the perturbation theory dependence of the integrand on the momentum transfer $u$ may be more complicate.

To calculate $T(N,x,\alpha)$ of Eq.(\ref{T}) for this model, we need to take into account that   
\begin{eqnarray}  
\label{int 1} \int_x^1 \frac{dy}{y} y^N \frac{1-y}{y} = \int_x^1~dy~y^{N-2} (1-y) = \frac{1}{N-1} - \frac{1}{N} + \frac{x^N}{N} - \frac{x^{N-1}}{N-1}, \\
\label{int 2} \int_x^1 \frac{dy}{y} y^N y(1-y) =   \frac{1}{N+1} - \frac{1}{N + 2} + \frac{x^{N+2}}{N+2} - \frac{x^{N+1}}{N+1}, \\
\label{int 3} \int_x^1 dy\frac{1-y^N}{1-y} = \psi(N+1) + C + \ln(1-x) + \frac{x^{N+1}}{N+1} + x^{N+2}\sum_{k=0}^{\infty}\frac{x^k}{N+k+2},
\end{eqnarray}
where $C$ is Euler-Mascheroni constant. Integral (\ref{int 3}) comes from the first term in the gluon splitting function  (\ref{almost realistic model})
which is defined as 
\begin{eqnarray} 
\int_0^1\frac{f(x)}{(1-x)_{+}}  = \int_0^1\frac{f(x) - f(1)}{1-x},  \hspace{1cm}
\int_x^1\frac{f(x)}{(1-x)_{+}}  = \int_x^1\frac{f(x) - f(1)}{1-x} + f(1)\ln(1-x).
\end{eqnarray}
This means that 
\begin{eqnarray}  
\int_x^1 dy\frac{y^N}{(1-y)̣̣_{+}} \equiv  \int_x^1 dy\frac{y^N-1}{1-y} + \ln(1-x)  = \no \\  
- \psi(N+1) - C  - \frac{x^{N+1}}{N+1} - x^{N+2}\sum_{k=0}^{\infty}\frac{x^k}{N+k+2},
\end{eqnarray}
This integral  generates harmonic numbers and generalizes them to the complex argument $z$, 
\begin{eqnarray} \label{Int-x}  
\int_x^1 dy\frac{1-y^z}{1-y} = -\int_{1-x}^0~du~\frac{1-(1-u)^z}{u} = \int_0^{1-x}~du~ \frac{1}{u} \le 1 - \sum_{k=0}^{\infty} \frac{(-z)_k}{k!} u^k \ri = \\  
- \int_0^{1-x}~du~ \sum_{k=1}^{\infty} \frac{(-z)_k}{k!} u^{k-1} = - \sum_{k=1}^{\infty} \frac{(-z)_k}{k \cdot k!} (1-x)^{k} \no
\end{eqnarray}
Here we use the well-known binomial expansion for an arbitrary complex power $z$ and $x \in [0,1].$
 \begin{eqnarray}  
(1-x)^z =  \sum_{k=0}^{\infty}\frac{\Gamma(-z+k)}{\Gamma(-z)}\frac{x^k}{k!} = \sum_{k=0}^{\infty} \frac{(-z)_k}{k!} x^k, 
\end{eqnarray}
in which $(a)_k = \Gamma(a+k)/\Gamma(a)$ stands for Pochhammer symbol. This formula may be derived by using Mellin-Barnes transformation \cite{Allendes:2012mr}.  
In particular case, when $x=0$ we obtain for integral (\ref{Int-x})
\begin{eqnarray} \label{Int-0}  
\int_0^1 dy\frac{1-y^z}{1-y}  = - \sum_{k=1}^{\infty} \frac{(-z)_k}{k \cdot k!}  = \psi(z+1) + C.
\end{eqnarray}
Also, another representation of Euler digamma function  necessary for future use  is 
\begin{eqnarray} \label{psi} 
\psi(z) = \sum_{k=1}^{\infty}\le \frac{1}{k} - \frac{1}{k+z-1} \ri - C
\end{eqnarray}
Taking into account that $\psi(n+1) = H_n - C,$ where $n$ is a natural number, integral (\ref{Int-0}) may be considered as an analytic continuation of harmonic numbers 
\begin{eqnarray}   
H_n = \sum_{k=1}^n \frac{1}{k}
\end{eqnarray}
to the complex plane $z.$ In such a case integral (\ref{Int-0}) is an analytic continuation of Euler integral 
\begin{eqnarray}   
\int_0^1 dy\frac{1-y^n}{1-y} = H_n. 
\end{eqnarray} 
According to Eq.(\ref{T}) and Eqs.(\ref{int 1}),(\ref{int 2}) and (\ref{int 3}) we have for $T(N,x,\alpha)$ with $P_{GG}$ (\ref{almost realistic model})  
\begin{eqnarray}  
\frac{1}{2C_2(G)}T\le N,x,\alpha\ri = - \psi(N+1) - C  + \frac{1}{N-1} - \frac{1}{N} +  \frac{1}{N+1} - \frac{1}{N + 2} \no \\
- \frac{x^{N-1}}{N-1} + \frac{x^N}{N} - \frac{2x^{N+1}}{N+1} + \frac{x^{N+2}}{N+2} - x^{N+2}\sum_{k=0}^{\infty}\frac{x^k}{N+k+2} =  \no \\
- \psi(N+1) - C  + \frac{1}{N-1} - \frac{1}{N} +  \frac{1}{N+1} - \frac{1}{N + 2} \no \\
- \frac{x^{N-1}}{N-1} + \frac{x^N}{N} - \frac{2x^{N+1}}{N+1} - x^{N+2}\sum_{k=1}^{\infty}\frac{x^k}{N+k+2}.
\end{eqnarray}
As we have mentioned, this Section is based on generalization of the solution for the toy-model considered in the previous Section. Thus, we should write for the anomalous dimension  
\begin{eqnarray}\label{gamma almost} 
\frac{1}{2C_2(G)}\gamma(N,\alpha) = \frac{1}{2C_2(G)}T(N,0,\alpha) =   - \psi(N+1) - C  + \frac{1}{N-1} - \frac{1}{N} +  \frac{1}{N+1} - \frac{1}{N + 2}
\end{eqnarray}

We note, that in this model the normalization condition 
\begin{eqnarray}
\gamma(1,\alpha)  = 1
\end{eqnarray}
cannot be maintained due to the pole in the complex plane at the point $N=1.$ In view of Eq.(\ref{psi}) we may re-write Eq.(\ref{gamma almost}) 
\begin{eqnarray}\label{gamma almost rewritten} 
\frac{1}{2C_2(G)}\gamma(N,\alpha) =  -\sum_{k=1}^{\infty}\le \frac{1}{k} - \frac{1}{N+k} \ri  + \frac{1}{N-1} - \frac{1}{N} +  \frac{1}{N+1} - \frac{1}{N + 2}
\end{eqnarray}

Thus, in analogy to the toy-model of the previous Section  Eq. (\ref{dglap T2}) may be rewritten in this case as 
\begin{eqnarray} 
\int_{a-i\infty}^{a+i\infty}~dN x^{-N} \phi_1(N) u^{\dis{\frac{\alpha}{2\pi}\gamma(N,\alpha)}} 
\left[\frac{x^{N-1}}{N-1} - \frac{x^N}{N} + \frac{2x^{N+1}}{N+1} + x^{N+2}\sum_{k=1}^{\infty}\frac{x^k}{N+k+2}\right]   =  \no \\
\frac{1}{x}\int_{a-i\infty}^{a+i\infty}~dN \frac{\phi_1(N)}{N-1} u^{\dis{\frac{\alpha}{2\pi}\gamma(N,\alpha)}}  -  
\int_{a-i\infty}^{a+i\infty}~dN \frac{\phi_1(N)}{N} u^{\dis{\frac{\alpha}{2\pi}\gamma(N,\alpha)}}  +  \no \\ 
+ 2x\int_{a-i\infty}^{a+i\infty}~dN \frac{\phi_1(N)}{N+1} u^{\dis{\frac{\alpha}{2\pi}\gamma(N,\alpha)}}  
+ \sum_{k=1}^{\infty}x^{k+2}\int_{a-i\infty}^{a+i\infty}~dN \frac{\phi_1(N)}{N+k+2} u^{\dis{\frac{\alpha}{2\pi}\gamma(N,\alpha)}} = 0.
\end{eqnarray}

We have obtained that some infinite series of the integer powers of $x$ must be zero. This means that the coefficient in front of each power 
is  zero, that is, the following identity must be fulfilled 
\begin{eqnarray} \label{Eq}
\int_{a-i\infty}^{a+i\infty}~dN \frac{\phi_1(N)}{N+k+2} u^{\dis{\frac{\alpha}{2\pi}\gamma(N,\alpha)}}  = 0,  ~~~ \forall k \in \mathbb{N} \cup \{-3,-2,-1\}.
\end{eqnarray}

In analogy to Eq. (\ref{dglap-4}) of the toy-model we obtain 
\begin{eqnarray} \label{dglap-5} 
0 = \int_{a-i\infty}^{a+i\infty}~dN \frac{\phi_1(N)u^{\dis{\frac{\alpha}{2\pi}\gamma(N,\alpha)}} }{N+k+2} =   \no\\
\sum_{j=0}^{\infty} \frac{1}{j!}\le\frac{\alpha}{2\pi}\ln{u}\ri^j \int_{a-i\infty}^{a+i\infty}~dN \frac{\phi_1(N)}{N+k+2}\le\gamma(N,\alpha)\ri^j   \Rightarrow \no\\  
\int_{a - i \infty}^{a + i \infty} d N \frac{\phi_1(N)}{N+k+2} \left[-\sum_{j=1}^{\infty}\le \frac{1}{j} - \frac{1}{N+j} \ri  + \frac{1}{N-1} - \frac{1}{N} +  \frac{1}{N+1} - \frac{1}{N + 2} \right]^m = 0,    
\end{eqnarray}
this equation is valid $\forall \{k,m\} \in \mathbb{N} \cup \{-3,-2,-1\}.$ When the coupling runs the self-consistency conditions will be a bit different (see Appendix \ref{Sccr}).

The solution for function $\phi_1(N)$ to Eq. (\ref{dglap-5}) is a linear combination of the terms like  
\begin{eqnarray} \label{terms} 
\prod_{j=1}^{n} (N-N_j)^{-\nu_j}
\end{eqnarray}
in which $\nu_j \in \mathbb{N} \cup \{0\}$ are arbitrary natural numbers or zero, $N_j$ belong to a set of arbitrary complex numbers such that ${\rm Re}~~ N_j < a,$
and at least one of the numbers $\nu_j$ should be nonzero. The requirement ${\rm Re}~~ N_j < a$ guarantees that all the poles of function  $\phi_1(N)$ appear to the left from
the vertical line of the contour in the complex plane $N.$ To fulfill Eq. (\ref{Eq}) by the terms of Eq. (\ref{terms}) we need to require that $a>1.$

To prove that a term like (\ref{terms}) gives a solution to Eq. (\ref{dglap-5}), we consider a simplified form of $\phi_1(N)$
\begin{eqnarray} \label{basic term} 
\phi_1(N) = \frac{1}{N-\lambda}, 
\end{eqnarray}
where $\lambda \in \mathbb{C}$ is an arbitrary complex number such that ${\rm Re}~~ \lambda < a.$ We may consider a term 
\begin{eqnarray} 
\frac{1}{(N-\lambda)(N+1)}  
\end{eqnarray}
According to the theory of transformation to Mellin moment  described in Section 2, all the poles should be situated to the left from the point $N =a$ in the complex plane of variable $N$ 
and the contour should be closed to the negative complex infinity because $x \in [0,1].$  If $a>1$ than 
\begin{eqnarray}\label{first term} 
\res_{N=\lambda}  \frac{1}{(N-\lambda)(N+1)} +  \res_{N=-1} \frac{1}{(N-\lambda)(N+1)} = 0. 
\end{eqnarray}
Let us consider another combination, 
\begin{eqnarray}  
\frac{1}{(N-\lambda)(N+1)^2} =  \frac{1}{N+1} \left[\frac{1}{N-\lambda}  -   \frac{1}{N+1}\right]   \frac{1}{\lambda+1}  =    \no\\
\frac{1}{\lambda+1}  \left[\frac{1}{(N-\lambda)(N+1)}  -  \frac{1}{(N+1)^2}\right],
\end{eqnarray}  
and we obtain again 
\begin{eqnarray} \label{second term} 
\res_{N=\lambda} \left[\frac{1}{(N-\lambda)(N+1)}  -  \frac{1}{(N+1)^2}\right] + \res_{N=-1} \left[\frac{1}{(N-\lambda)(N+1)}  -  \frac{1}{(N+1)^2}\right] = 0. 
\end{eqnarray}
The terms of second degree or higher do not contribute into residue calculus due to Cauchy formula and the first term does not contribute due to Eq. (\ref{first term}).
The third type of terms, which we consider in this proof, is 
\begin{eqnarray}   
\frac{1}{(N-\lambda)(N+1)(N+2)} =  \frac{1}{N-\lambda}\left[\frac{1}{N+1} - \frac{1}{N+2}\right].
\end{eqnarray}
Such a representation means that 
\begin{eqnarray} \label{third term} 
\res_{N=\lambda} \frac{1}{(N-\lambda)(N+1)(N+2)} +  \res_{N=-1} \frac{1}{(N-\lambda)(N+1)(N+2)} + \no\\
+ ~~~  \res_{N=-2} \frac{1}{(N-\lambda)(N+1)(N+2)} = 0.
\end{eqnarray}
At the end of this proof, we observe that any term of type like in Eq (\ref{terms}) may be decomposed in a finite sum of terms (\ref{basic term}) or their natural powers. 
Formulas (\ref{first term}), (\ref{second term}) and (\ref{third term}) show that  the term (\ref{basic term}) is a solution to  Eq. (\ref{dglap-5}) if $a > 1.$

Thus, any linear combination of the terms like (\ref{terms}) can be used for function $\phi_1(N).$ To show how the residue calculus works for this solution, we take again 
the simplest case
\begin{eqnarray}   
\phi_1(N) = \frac{1}{N+1}, 
\end{eqnarray}
which has been used in the previous Section for the toy-model and has appeared to be successful in reproducing the Bessel-like behaviour of unintegrated gluon distribution
$\phi(x,u)$ reviewed in Re.\cite{Salam:1999cn}. 

The result of calculation for the first two orders of expansion in terms of powers $\dis{\frac{\alpha}{2\pi}\ln{u}}$ is 
\begin{eqnarray}  
\phi(x,u) = \int_{a-i\infty}^{a+i\infty}~dN \phi_1(N)u^{\dis{\frac{\alpha}{2\pi}\gamma(N,\alpha)}}x^{-N}  =   
\int_{a-i\infty}^{a+i\infty}~dN \frac{x^{-N}}{N+1} u^{\dis{\frac{\alpha}{2\pi}\gamma(N,\alpha)}} =  \no\\
\sum_{j=0}^{\infty} \frac{1}{j!}\le\frac{\alpha}{2\pi}\ln{u}\ri^j \int_{a-i\infty}^{a+i\infty}~dN \frac{x^{-N}}{N+1}\le\gamma(N,\alpha)\ri^j \no = \\
x + \frac{C_2(G)\alpha}{\pi}\ln{u}\le x\ln{(1-x)} -2x\ln{x}  + \frac{1-x^2}{2x}  - (1-x^2) \ri  + o\le\frac{\alpha}{2\pi}\ln{u}\ri  \label{result}
\end{eqnarray}
The integrals may be taken by Cauchy formula in each power of $\ln{u}.$ The integral in front of the first power  of $\dis{\frac{\alpha}{2\pi}\ln{u}}$ is
\begin{eqnarray} 
\int_{a - i \infty}^{a + i \infty} d N \frac{x^{-N}}{N+1} \left[-\sum_{j=1}^{\infty}\le \frac{1}{j} - \frac{1}{N+j} \ri  + \frac{1}{N-1} - \frac{1}{N} +  \frac{1}{N+1} - \frac{1}{N + 2} \right] = \no  \\
x\le\ln{(1-x)} -  \ln{x}\ri  + \frac{1}{2}\le \frac{1}{x} - x \ri - (1-x) - (x -x^2) - x\ln{x} =  \no \\
-2x\ln{x} + x\ln{(1-x)}   + \frac{1-x^2}{2x}  - (1-x^2) 
\end{eqnarray}
Here we take into account the integrals 
\begin{eqnarray} 
\int_{a - i \infty}^{a + i \infty} d N \frac{x^{-N}}{(N+1)(N-1)} = \frac{1}{2}\int_{a - i \infty}^{a + i \infty} d N x^{-N}\left[ \frac{1}{N-1} - \frac{1}{N+1} \right] &=& \frac{1}{2}\le \frac{1}{x} - x \ri, \no\\
\int_{a - i \infty}^{a + i \infty} d N \frac{x^{-N}}{(N+1)N} &=& 1-x, \no \\
\int_{a - i \infty}^{a + i \infty} d N \frac{x^{-N}}{(N+1)(N+2)} &=& x -x^2, \no \\
\int_{a - i \infty}^{a + i \infty} d N \frac{x^{-N}}{(N+1)^2} &=&  -x\ln{x} 
\end{eqnarray}
and the integral  
\begin{eqnarray} 
\int_{a - i \infty}^{a + i \infty} d N \frac{x^{-N}}{N+1} \sum_{j=1}^{\infty}\le \frac{1}{j} - \frac{1}{N+j} \ri  = \no \\
\int_{a - i \infty}^{a + i \infty} d N \frac{x^{-N}}{N+1} \left[1 -  \frac{1}{N+1}  + \sum_{j=2}^{\infty}\le \frac{1}{j} - \frac{1}{N+j} \ri\right] =  \no \\
\int_{a - i \infty}^{a + i \infty} d N x^{-N} \left[\frac{1}{N+1} -  \frac{1}{(N+1)^2}  +  \frac{1}{N+1} \sum_{j=2}^{\infty}\le \frac{1}{j}  - \frac{1}{j-1}\ri \right. 
+ \left.  \sum_{j=2}^{\infty} \frac{1}{(j-1)(N+j)} \right] = \no \\
x\ln{x} + \sum_{j=2}^{\infty}  \frac{x^j}{j-1} = x(\ln{x} - \ln{(1-x)}). 
\end{eqnarray}

As we may see in Eq.(\ref{result}) there are singularities at the points $x=0$ and $x=1$ at the first order of the expansion in terms of $\le{\alpha}/{2\pi}\ri\ln{u}.$
We sum the leading terms of these singularities and show that the singularity at $x=0$ survives while the singularity at $x=1$ disappears. First, we treat 
the singularity at the point $x=0.$ It is produced by the residue at $N=1.$ The most singular contribution is produced by the natural powers of 
$1/(N-1)$ in each term of the expansion in Eq.(\ref{result}) because in addition to $1/x$ we will obtain factor $\ln{(1/x)}$ in the maximal power. Thus, in the vicinity of the point $x=0$ we may write 
\begin{eqnarray*}  
\phi(x,u) = \sum_{j=0}^{\infty} \frac{1}{j!}\le\frac{\alpha}{2\pi}\ln{u}\ri^j \int_{a-i\infty}^{a+i\infty}~dN \frac{x^{-N}}{N+1}\le\gamma(N,\alpha)\ri^j \sim \no \\
\sim  \sum_{j=0}^{\infty} \frac{1}{j!}\le\frac{\alpha}{2\pi}\ln{u}\ri^j \int_{a-i\infty}^{a+i\infty}~dN \frac{x^{-N}}{N+1}\le\frac{2C_2(G)}{N -1}\ri^j  \no \\
\sim  x + \frac{1}{2}\sum_{j=1}^{\infty} \frac{1}{j!}\le\frac{C_2(G)\alpha}{\pi}\ln{u}\ri^j \int_{a-i\infty}^{a+i\infty}~dN \frac{x^{-N}}{(N-1)^j} = \no \\
x + \frac{1}{2x}\sum_{j=1}^{\infty} \frac{1}{j!}\le\frac{C_2(G)\alpha}{\pi}\ln{u}\ri^j \frac{\le -\ln{x}\ri^{j-1}}{(j-1)!} \leqslant 
x + \frac{1}{2x}\sum_{j=1}^{\infty} \frac{1}{j!}\le\frac{C_2(G)\alpha}{\pi}\ln{u}\ri^j \le \ln{\frac{1}{x}}\ri^{j-1} = \no \\
\end{eqnarray*}
\begin{eqnarray}  
x + \frac{1}{2x\dis{\ln(1/x)}}\le e^{\dis{\frac{C_2(G)\alpha}{\pi}\ln{u} \ln{\frac{1}{x}} }} -1\ri = x + \frac{1}{2x\dis{\ln(1/x)}}\le \le\frac{1}{x} \ri^{\dis{\frac{C_2(G)\alpha}{\pi}\ln{u}  }} -1\ri \no\\
\sim \frac{1}{2x\dis{\ln(1/x)}}\le \le\frac{1}{x} \ri^{\dis{\frac{C_2(G)\alpha}{\pi}\ln{u}  }} -1\ri \sim \frac{1}{2\dis{\ln(1/x)}} \le\frac{1}{x} \ri^{\dis{1+\frac{C_2(G)\alpha}{\pi}\ln{u}  }} 
\end{eqnarray}
This equation gives by itself an upper bound on unintegrated gluon distribution $\phi(x,u)$ in the vicinity of the point $x=0.$ The upper bound is a singular function at the limit $x \rightarrow 0.$ To be sure that  
$\phi(x,u)$ is a singular function we need to consider a lower bound for it in the vicinity of the point   $x=0,$ 
\begin{eqnarray}  
\phi(x,u) = \sum_{j=0}^{\infty} \frac{1}{j!}\le\frac{\alpha}{2\pi}\ln{u}\ri^j \int_{a-i\infty}^{a+i\infty}~dN \frac{x^{-N}}{N+1}\le\gamma(N,\alpha)\ri^j \sim \no \\
\sim   x + \frac{1}{2} \sum_{j=1}^{\infty} \frac{1}{j!} \left( \frac{ C_2(G)\alpha}{\pi} \ln u \right)^j  \int_{a - i \infty}^{a+i \infty} dN \frac{x^{-N}}{(N-1)^j} = \no \\
x + \frac{1}{2 x} \sum_{j=1}^{\infty} \frac{1}{j!} \left( \frac{C_2(G) \alpha}{\pi} \ln u \right)^j \left( \ln \frac{1}{x} \right)^{j-1} \frac{1}{(j-1)!} >
x + \frac{1}{2 x} \frac{1}{\ln(1/x)} \sum_{j=1}^{\infty} \frac{1}{(j!)^2} \left( \frac{ C_2(G)\alpha}{\pi} \ln u  \ln \frac{1}{x} \right)^j  \no\\
 =  x + \frac{1}{2 x}  \frac{1}{\ln(1/x)} \left[ I_0(K z) - 1 \right]{\bigg |}_{z=\sqrt{\ln(1/x)},  K= \sqrt{(4 C_2(G)\alpha/\pi) \ln u}}
\sim  \frac{1}{2 x} {\cal G}(z, K) \ ,
\label{phix0}
\end{eqnarray}
where, when using the asymptotic behavior of the modified Bessel function $I_0(z)$, we have
\begin{eqnarray}  
{\cal G}(z, K) & \equiv & \frac{1}{z^2} \left[ I_0(Kz) - 1 \right] 
\nonumber\\
&= &\frac{1}{z^2} \left[ \frac{\exp(K z)}{\sqrt{2 \pi K z}}\left(1 + {\cal O}(1/z) \right) - 1 \right]
\label{calG}
\end{eqnarray}
and $z\equiv \sqrt{\ln(1/x)}$ and $K= \sqrt{(4C_2(G)\alpha/\pi) \ln u}$. 

If we assume $K>0$  ($u>1$), then the function ${\cal G}(z, K)$ has the behavior
\begin{eqnarray} 
{\cal G}(z,K) \to +\infty \qquad {\rm when \ } z \to + \infty.
\label{calGasymp}
\end{eqnarray}
Therefore, we have, by Eqs.~(\ref{phix0}) and (\ref{calGasymp})
\begin{eqnarray} 
\phi(x,u) \to +\infty \qquad {\rm when \ } x \to +0.
\end{eqnarray}
We conclude that a lower bound for the unintegrated gluon distribution in the vicinity of the point $x=0$ is determined by the modified Bessel function and it is singular 
in the small $x$ region.

On the contrary, the singularity at the point $x=1$ disappears.  
We may conclude for the considerations presented in the previous paragraphs of this Section that the most singular contribution is the biggest power of $\ln{(1-x)}.$ This may come only from powers of 
$H(N) = \psi(N+1) + C$  Harmonic number function in Eq. (\ref{result}). If we consider integral 
\begin{eqnarray} \label{squ} 
\int_{a - i \infty}^{a + i \infty} d N \frac{x^{-N}}{N+1} \left[\sum_{j=1}^{\infty}\le \frac{1}{j} - \frac{1}{N+j} \ri \right]^2,
\end{eqnarray}
we conclude 
by considering carefully the singularity structure of the Harmonic number function $H(N) = \psi(N+1) + C$ in the complex plane, applying repeatedly the Cauchy theorem, 
and then using the identities
\begin{eqnarray}  
\sum_{j=1}^{\infty}H_jx^j = -\frac{\ln(1-x)}{1-x}, \\    
\sum_{j=1}^{\infty}H_j\frac{x^j}{j} = \frac{1}{2}\ln^2(1-x) + {\rm Li}̣_2(x), \\   
\sum_{j=1}^{\infty}\frac{x^j}{j(j+1)} = \frac{(1-x)\ln(1-x)}{x} + 1
\end{eqnarray}
that in the vicinity of the point $x=1$ the result for integral (\ref{squ})  has the following asymptotic behaviour 
\begin{eqnarray} 
\int_{a - i \infty}^{a + i \infty} d N \frac{x^{-N}}{N+1} \left[\sum_{j=1}^{\infty}\le \frac{1}{j} - \frac{1}{N+j} \ri \right]^2 \sim  x\ln^2{(1-x)}.
\end{eqnarray}
The same is true for the higher power of the $\psi(N+1)$ function in the integrand of  Eq. (\ref{result}). Thus, at the vicinity of the point $x=1$ we may write 
\begin{eqnarray} \label{last} 
\phi(x,u) = \sum_{j=0}^{\infty} \frac{1}{j!}\le\frac{\alpha}{2\pi}\ln{u}\ri^j \int_{a-i\infty}^{a+i\infty}~dN \frac{x^{-N}}{N+1}\le\gamma(N,\alpha)\ri^j \sim \no \\
\sim  \sum_{j=0}^{\infty} \frac{1}{j!}\le\frac{\alpha}{2\pi}\ln{u}\ri^j \int_{a-i\infty}^{a+i\infty}~dN \frac{x^{-N}}{N+1}\le -2 C_2(G)\le \psi(N+1) + C\ri \ri^j \no \\
\sim  x\sum_{j=0}^{\infty} \frac{1}{j!}\le\frac{C_2(G)\alpha}{\pi}\ln{u}\ri^j \le\ln{(1-x)}\ri^j =  x e^{\dis{\frac{C_2(G)\alpha}{\pi}\ln{u}~\ln{(1-x)} }} = 
x\le 1-x \ri^{\dis{\frac{C_2(G)\alpha}{\pi}\ln{u}  }} 
\end{eqnarray}
We observe that the highest singularities at the point $x=1$ disappear after summing the leading singularities up. This is in agreement with Eq.(\ref{Eq}).  Indeed, 
$\phi(1,u)$ looks like Eq.(\ref{Eq}) without the denominator in the integrand.

\section{Conclusion}

In the present article we have found a way to solve DGLAP integro-differential equation analytically. The method we propose is simple and is based on the fact 
that integrals of the splitting functions in the range from $0$ till $x$ (where $x$ is  Bjorken variable) are proportional to $x^N$ where $N$ is the complex variable of the 
Mellin moment $\phi(N,Q^2/\mu^2)$ of the unintegrated gluon distribution $\phi(x,Q^2/\mu^2)$, cf.~Eq.~(\ref{conclusion-11}).
Due to cancellation of this power $x^N$ with the power $x^{-N}$ which stands in the inverse integral transformation, cf.~Eqs.~(\ref{conclusion-21})-(\ref{dglap T2}),
we obtain an expansion in terms of integer powers of $x$ from which we may conclude that the coefficient in front of each integer power of $x$ must be zero. These requirements give us 
a set of integrals involving Mellin moment $\phi(N,Q^2/\mu^2)$  of unintegrated gluon distribution  $\phi(x,Q^2/\mu^2)$  which must be equal to zero simultaneously,  cf.~Eqs.~(\ref{dglap x degree}) and (\ref{Eq}).
We have found a way to solve these integral restrictions analytically by making use of Cauchy formula. The method we have found may have a wide spectrum of applications 
in science and technology.

We have considered a simple toy-model of DIS processes and found an analytical solution for the DGLAP equation in this toy-model. A simplified splitting function  
(\ref{toy model}) was used as an input.  The Mellin moment $\phi(N,Q^2/\mu^2)$ of  unintegrated gluon distribution $\phi(x,Q^2/\mu^2)$ appears to be a linear combination of the chosen terms.  
The infinite set of constants $c_j$ which are coefficients in front of these chosen terms  remains unfixed in this toy-model.  
The solution is parametrized by them.  It could be that they are fixed  if we consider DGLAP IDE together with BFKL IDE.  
However, we have shown in this article  that the corresponding DGLAP IDE by itself contains enough information to represent the chosen unintegrated gluon distribution $\phi(x,Q^2/\mu^2)$ 
in this toy-model in the form of expansion  in terms of $\ln{Q^2/\mu^2}$ and $\ln{1/x}$ shown in Eq.(\ref{general case}).

When we choose only one simplest term from all the possible terms, we obtain a Bessel-like behaviour for  unintegrated gluon distribution 
$\phi(x,Q^2/\mu^2).$  Such a behaviour of $\phi(x,Q^2/\mu^2)$  has been obtained in Ref.\cite{Salam:1999cn} by summing ladder diagrams 
in an estimative way for the realistic splitting function $P_{GG}(z)$ in a pure gluonic Chromodynamics. We have shown that such a behaviour corresponds to 
the selection of this simplest term from all the possible terms for the Mellin moment $\phi(N,Q^2/\mu^2)$ of $\phi(x,Q^2/\mu^2)$ in our toy-model with the simplified 
splitting function.

Situation becomes more complicated for the realistic one-loop splitting function  $P_{GG}(z).$ The number of the possible terms for the Mellin moment $\phi(N,Q^2/\mu^2)$ is infinite too, however
more rich structure of the splitting function produces more complicate anomalous dimension for unintegrated gluon distribution. As the result, the distribution  $\phi(x,Q^2/\mu^2)$
looks more complicated than for the toy-model.  This happens even in the case  when the same simplest term like in the toy model is selected of all the possible terms 
for Mellin moment $\phi(N,Q^2/\mu^2).$   Making complex integrals by use of Cauchy formula for the selected simple term of the Mellin moment $\phi(N,Q^2/\mu^2),$ we obtain 
distribution $\phi(x,Q^2/\mu^2)$ as an expansion in powers of $\alpha\ln{Q^2/\mu^2}.$

The summation of this expansion in powers of $\alpha\ln{Q^2/\mu^2}$ looks difficult in this realistic case. However, the second term of the expansion shows singularities at the points 
$x=0$ and $x=1$ whose origin in the complex plane of variable $N$ may be detected and the corresponding terms responsible for these singularities may be analysed. 
These singular terms at the  points $x=0$ and $x=1$ may be summed up in all the orders of the expansion in powers of  $\alpha\ln{Q^2/\mu^2}.$ After summing up these singularities 
at the point $x=1,$ they disappear and the behaviour of unintegrated gluon distribution $\phi(x,Q^2/\mu^2)$ becomes smooth with respect to variable $x$ in the vicinity of the point $x=1.$ 
However, the sum of the singular terms at the point $x=0$ taken to all orders of  $\alpha\ln{Q^2/\mu^2}$ remains singular with respect to $x$ at the point $x=0.$
The result of summation shows the Bessel-like behaviour in the vicinity of $x=0$ which is similar to the behavior of unintegrated gluon distribution obtained 
in Ref. \cite{Salam:1999cn} by summing ladder diagrams or by calculating integrals via saddle-point method.

We found in this paper a large set of solutions to the DGLAP equation without using any other information from any additional equation. In particular, we did not use any information from 
BFKL equation. This may be considered as an alternative way to the 
approach of Refs.\cite{Ball:1999sh}-\cite{Ball:2005mj} where BFKL IDE has been widely used. We have shown that this integro-differential equation has infinitely many solutions
for any given kernel $P(z)$ by itself if we do not provide any boundary condition for unknown parton distributions.

\subsection*{Acknowledgments}

The work of G.A. was supported in part by the joint DAAD-Conicyt (Chile) scholarship and by Fondecyt (Chile) Grant No.\ 1121030. 
The work of G.C. was supported in part by Fondecyt (Chile) Grant No.~1220095.
The work of B.A.K. was supported in part by the German Science Foundation (DFG) within the Collaborative Research Center SFB 676 ``Particles, Strings and the
Early Universe'' and by the German Federal Ministry for Education and Research
(BMBF) through Grant No.\ 05H12GUE.
The work of I.K. was supported in part 
by Fondecyt (Chile) Grants Nos. 1040368, 1050512 and 1121030, by DIUBB (Chile) Grant Nos.  125009,  GI 153209/C  and GI 152606/VC.  
Also, the work of I.K. is supported by Universidad del B\'\i o-B\'\i o and Ministerio de Educacion (Chile) within Project No.\ MECESUP UBB0704-PD018.
He is grateful to the Physics Faculty of Bielefeld University for accepting
him as a visiting scientist and for the kind hospitality and the excellent
working conditions during his stay in Bielefeld. 
The work of I.P.F. was supported in part by Fondecyt (Chile) Grant No.\ 1121030 and by Beca Conicyt (Chile) via Master fellowship CONICYT-PCHA/Magister Nacional/2013-22131319. 
A part of these results was presented in the talk of I.K. at LXXXIV Encuentro Anual Sociedad de Matem\'atica de Chile, Puc\'on, Chile, Novembre 26 - 28, 2015.
He is grateful to V\'ictor H. Cort\'es for inviting him to give a talk at Section ``Functional Analysis and Applications'' of this annual scientific meeting.

\appendix 

\section{Running coupling case} \label{Run}

In the case when the gauge coupling  runs, that is the case of QCD,  the first order differential DGLAP equation (\ref{dglap approx}) in the small $x$ limit for the Mellin moment  $G\le N,u\ri$ of  the  dominant PDF
\begin{eqnarray} \label{Run-1} 
u\frac{d}{du} G(N, u) = \frac{\alpha(u)}{2\pi}\gamma(N,\alpha(u))G(N,u) \Rightarrow  \frac{d}{d\alpha} G(N, u(\alpha)) = \frac{\alpha}{2\pi}\frac{\gamma(N,\alpha)}{\beta(\alpha)}G(N,u(\alpha)),
\end{eqnarray} 
where the coupling  $\alpha$ has been chosen as a variable in a usual way instead of the scale $u$ for this equation, $\beta(\alpha) = u~d\alpha(u)/du.$ Then,    
\begin{eqnarray}\label{Run-2} 
G(N, u(\alpha)) =  G(N,u(\alpha_0))\exp{F(N,\alpha)}, ~~~{\rm  where}~~~~  \frac{d}{d\alpha} F(N,\alpha) = \frac{\alpha}{2\pi}\frac{\gamma(N,\alpha)}{\beta(\alpha)}
\end{eqnarray} 
Such a change of variable  requires that $\alpha(u)$ is a monotonic function.  This is true in the perturbation high energy QCD \cite{Cvetic:2011vy,Ayala:2017tco}. 
Here  $G\le N,u(\alpha_0)\ri =  G\le N,1\ri$ is a Mellin moment of the shape function at the scale $Q^2 = \mu^2,$ that is, at $u=1,$ $\alpha_0 = \alpha(1).$  According to our notation, 
$G\le N,1\ri$ appears to be the Mellin moment of $ G(x,1) .$   This function  $G(x,1)$  should be parametrized. We mentioned in the Introduction of Ref.\cite{Kondrashuk:2019cwi} various known parametrizations  of the PDF shapes 
at some fixed momentum transfers for the case of QCD.

\section{DGLAP for unintegrated PDFs: running coupling case} \label{Updr}

If the coupling runs, what is the case of QCD, a construction based on the unintegrated PDF (\ref{UGD}) which satisfies DGLAP equation  (\ref{dglap approx}) is different from 
$\dis{Q^2\varphi \le N,Q^2/\mu^2\ri}$ which we had seen in Section \ref{Updf} dedicated to the frozen coupling constant.

Let us write again Eq. (\ref{dglap approx}) in the form of (\ref{UGD-2}) but in this case when the coupling runs
\begin{eqnarray*} 
Q^2\frac{d}{dQ^2} \int_0^{Q^2}dk_\perp^2 \varphi\le N,\frac{k_\perp^2}{\mu^2}\ri = \frac{\alpha\le Q^2/\mu^2\ri}{2\pi}\gamma(N,\alpha\le Q^2/\mu^2\ri ) \int_0^{Q^2}dk_\perp^2 \varphi\le N,\frac{k_\perp^2}{\mu^2}\ri, 
\end{eqnarray*}
and re-write it in the following form 
\begin{eqnarray*} 
\frac{2\pi~~  Q^2 \varphi\le N, Q^2/\mu^2 \ri}{\alpha\le Q^2/\mu^2\ri\gamma(N,\alpha\le Q^2/\mu^2\ri ) } =  \int_0^{Q^2}dk_\perp^2 \varphi\le N,\frac{k_\perp^2}{\mu^2}\ri, 
\end{eqnarray*}
this means the following first order differential equation is valid 
\begin{eqnarray*} 
Q^2\frac{d}{dQ^2} \frac{2\pi~~  Q^2 \varphi\le N, Q^2/\mu^2 \ri}{\alpha\le Q^2/\mu^2\ri\gamma(N,\alpha\le Q^2/\mu^2\ri ) } =  Q^2 \varphi\le N, Q^2/\mu^2 \ri \\
=  \frac{\alpha\le Q^2/\mu^2\ri}{2\pi}\gamma(N,\alpha\le Q^2/\mu^2\ri  )\frac{2\pi~~  Q^2 \varphi\le N, Q^2/\mu^2\ri}{\alpha\le Q^2/\mu^2\ri\gamma(N,\alpha\le Q^2/\mu^2\ri ) }.
\end{eqnarray*}
Thus, the dimensionless combination  
\begin{eqnarray} \label{Updr-1} 
\phi(N, u) =  \phi(N, Q^2/\mu^2) = \frac{2\pi~~  Q^2 \varphi\le N, Q^2/\mu^2 \ri}{\alpha\le Q^2/\mu^2\ri\gamma(N,\alpha\le Q^2/\mu^2\ri ) } 
\end{eqnarray}
of the unintegrated  dominant PDF  $\varphi\le N, Q^2/\mu^2 \ri$ from Eq. (\ref{UGD}), of the momentum transfer $Q^2,$ of the running coupling     $\alpha\le Q^2/\mu^2\ri$ and of the 
anomalous dimension $\gamma(N,\alpha\le Q^2/\mu^2\ri)$ satisfies the same first order differential equation  (\ref{dglap approx}) as well as the moment of its integrated  dominant PDF does.

Doing the inverse Mellin transformation from the Mellin moment $\phi(N, u)$ to the dimensionless dominant PDF $\phi(x, u)$ with respect to complex variable $N,$
we obtain a set of equations almost identical to the set of Eqs. (\ref{dglap phi classic}-\ref{dglap gamma}), the only difference is that the coupling depends on the momentum transfer $Q^2,$
\begin{eqnarray} \label{Updr-2}
u \frac{d}{d u} \phi \le x,u\ri = \frac{\alpha(u)}{2\pi}\int_x^1\frac{d y}{y}\phi \le y, u\ri P_{GG}\le \frac{x}{y},\alpha(u)\ri,    \label{dglap phi classic 2} \\
u \frac{d}{d u} \phi \le N, u\ri   = \frac{\alpha(u)}{2\pi}\gamma(N,\alpha(u)) \phi\le N, u\ri, \label{dglap phi N u meq 2} \\
\phi \le N,u\ri = \int_0^{1}dx~x^{N-1}\phi\le x, u\ri,  \label{dglap phi N phi x u 2} \\
\gamma(N,\alpha(u)) = \int_0^1 dx ~x^{N-1} P_{GG}\left(x,\alpha(u) \right), 
\end{eqnarray}
Here we have $\phi \le N,u\ri$ defined in terms of Mellin moment of unintegrated dominant PDF  $\varphi\le N, u \ri$ by the relation  (\ref{Updr-1}).  
Eq. (\ref{dglap phi classic 2}) repeats exactly Eq.(\ref{dglap approx}) for its integrated  dominant PDF,  this means the solution to Eq. (\ref{dglap phi N u meq 2}) repeats 
exactly the solution to Eq. (\ref{Run-1}),
\begin{eqnarray} \label{Updr-3}
u\frac{d}{du} \phi(N, u) = \frac{\alpha(u)}{2\pi}\gamma(N,\alpha(u))\phi(N,u) \Rightarrow  \frac{d}{d\alpha} \phi(N, u(\alpha)) = \frac{\alpha}{2\pi}\frac{\gamma(N,\alpha)}{\beta(\alpha)}\phi(N,u(\alpha)),
\end{eqnarray} 
where the change of the differentiation variable from the momentum transfer $u$ to the coupling  $\alpha$  is done. Solving the differential equation above, we come to 
\begin{eqnarray} \label{Updr-4} 
\phi(N, u(\alpha)) =  \phi(N,u(\alpha_0))\exp{F(N,\alpha)}, ~~~{\rm  where}~~~~  \frac{d}{d\alpha} F(N,\alpha) = \frac{\alpha}{2\pi}\frac{\gamma(N,\alpha)}{\beta(\alpha)}.
\end{eqnarray} 
We may do the same comments, that we have done in Appendix \ref{Run} for the solution to the differential equation for the Mellin moment $G(N,u)$ of the integrated dominant PDF, in the QCD case, 
that is, in the case when the coupling runs. Namely, the change of variables from the momentum transfer $u$ to the coupling   $\alpha$
supposes one-to-one correspondence between $u$ and $\alpha.$ This happens at least in the penetrative high energy QCD \cite{Cvetic:2011vy,Ayala:2017tco}. 
Here  $\phi\le N,u(\alpha_0)\ri =  \phi\le N,1\ri \equiv \phi_1(N) $ is the Mellin moment of the shape function at the scale $Q^2 = \mu^2,$ that is, at $u=1,$ $\alpha_0 = \alpha(1).$ 
The moment $\phi(N,1)$ of the shape function $\phi(x,1)$
for the unintegrated dominant PDF may be obtained from the solution (\ref{Run-2}) to the integrated dominant PDF $G(N,u)$ and the parametrization 
of the shape function $G(x,1).$  We have written in Ref.\cite{Kondrashuk:2019cwi} about the parameterizations which are frequently used for the shape function. It may be proven from 
Eqs. (\ref{Run-1}) and (\ref{UGD}) that 
\begin{eqnarray*} 
\phi\le N,1\ri \equiv   \phi_1\le N \ri =  G(N,1).
\end{eqnarray*}
This means, the shape function $\phi\le N,1\ri$ of the unintegrated dominant PDF coincides with the shape function  $G(N,1)$ of the integrated dominant PDF, and they are parametrized identically.

\section{Self-consistent shape function for the running coupling}  \label{Sccr}

The DGLAP IDE (\ref{dglap phi classic 2}) has a solution in the form  of Eq. (\ref{Updr-4})  for the Mellin $N$-moment of the unintegrated dominant PDF $\phi(N,u).$ This solution does not restrict the form  of the function $\phi_1(N).$  
The reason is that when we do the  integration over variable $x$ on both sides of IDE (\ref{dglap phi classic 2}), we are averaging the information about $x$ in the unintegrated  dominant PDF $\phi(x,u).$
After this averaging we obtain a differential equation for the Mellin moments like Eqs. (\ref{Updr-3}) and (\ref{dglap phi N u meq 2}).

However,  as in the case of fixed $\alpha$ we may look at DGLAP IDE at a different angle and substitute the inverse transformation (\ref{MMT Cauchy}) in DGLAP IDE (\ref{dglap phi classic 2}) for the unintegrated  dominant PDF  
$\phi(x,u).$ Such a strategy  should give restrictions on the function  $\phi_1(N),$ because we use pointwise information.  Indeed, by doing this we obtain 
\begin{eqnarray} \label{Sccr-1} 
u\frac{d}{d u} \phi \le x,u\ri = \frac{\alpha(u)}{2\pi}\int_x^1\frac{d y}{y}\phi \le y, u\ri P_{GG}\le \frac{x}{y},\alpha(u)\ri     \no\\
 \Rightarrow  u\frac{d}{d u}\int_{a-i\infty}^{a+i\infty}~dN x^{-N} \phi(N,u) = \frac{\alpha(u)}{2\pi}\int_x^1\frac{d y}{y}\int_{a-i\infty}^{a+i\infty}~dN y^{-N} \phi(N,u) P_{GG}\le \frac{x}{y},\alpha(u)\ri   \no\\
 \Rightarrow  \int_{a-i\infty}^{a+i\infty}~dN x^{-N} \phi_1(N) e^{\dis{F(N,\alpha)}}  \gamma(N,\alpha)   \no\\
= \int_x^1\frac{d y}{y}\int_{a-i\infty}^{a+i\infty}~dN y^{-N}  \phi_1(N) e^{\dis{F(N,\alpha)}} P_{GG}\le \frac{x}{y},\alpha\ri   \no\\
\Rightarrow \int_{a-i\infty}^{a+i\infty}~dN x^{-N} \phi_1(N)e^{F(N,\alpha)}\left[\gamma(N,\alpha) - x^N\int_x^1\frac{d y}{y} y^{-N}  P_{GG}\le \frac{x}{y},\alpha\ri \right] = 0.
\end{eqnarray} 
The integral in the bracket may be transformed to    
\begin{eqnarray} \label{Sccr-2} 
\int_x^1 \frac{dy}{y} ~y^{-N} P_{GG}\le \frac{x}{y},\alpha\ri = \int_1^{1/x} \frac{dy}{y} ~y^{N} P_{GG}(xy,\alpha) 
= x^{-N}\int_x^1 \frac{dy}{y} y^{N} P_{GG}(y,\alpha).
\end{eqnarray}
The DGLAP IDE may be written in such a form  
\begin{eqnarray} \label{Sccr-3} 
\int_{a-i\infty}^{a+i\infty}~dN x^{-N} \phi_1(N)e^{F(N,\alpha)}\left[\gamma(N,\alpha) - \int_x^1 \frac{dy}{y} y^{N} P_{GG}(y,\alpha)\right] \no \\
= \int_{a-i\infty}^{a+i\infty}~dN x^{-N} \phi_1(N)e^{F(N,\alpha)}\int_0^x \frac{dy}{y} y^{N} P_{GG}(y,\alpha) = 0.
\end{eqnarray}
The main idea to get self-consistency condition is the contour integral should be put to zero in front of each power of expansion in terms of $x$ on the right hand 
side of Eq. (\ref{Sccr-3})  for the same contour.

\end{document}